\documentclass[journal]{IEEEtran}
\usepackage{cite}

\usepackage{multirow}
\usepackage{float}

\ifCLASSINFOpdf
  \usepackage[pdftex]{graphicx}
  \usepackage{subfigure}
\else
\fi
\usepackage{amsmath}
\usepackage{algorithm}
\begin{document}
%
\title{\huge An 8-bit In Resistive Memory Computing Core with Regulated Passive Neuron and Bit Line Weight Mapping}
%
%
%

\author{Yewei~Zhang,~\IEEEmembership{Student~Member,~IEEE},
        Kejie~Huang,~\IEEEmembership{Senior~Member,~IEEE},
        Rui~Xiao,~\IEEEmembership{Student~Member,~IEEE},
        and~Haibin~Shen
\thanks{Authors: K. Huang and H. Shen are with the College of Information Science \& Electronic Engineering, Zhejiang University, 38 Zheda Road, Hangzhou, China, 310027, and also with Zhejiang Lab, Building 10, China Artificial Intelligence Town,	1818 Wenyi West Road, Hangzhou City, Zhejiang Province, China, email: huangkejie@zju.edu.cn; shen\_hb@zju.edu.cn	Y. Zhang and R. Xiao are with the College of Information Science \&	Electronic Engineering, Zhejiang University, 38 Zheda Road, Hangzhou, China, 310027, email: yeweizhang@zju.edu.cn; xiaor@zju.edu.cn}
\thanks{}
\thanks{}}

%
%

\markboth{}%
{Shell \MakeLowercase{\textit{et al.}}: Bare Demo of IEEEtran.cls for IEEE Journals}
%



\maketitle

\begin{abstract}
The rapid development of Artificial Intelligence (AI) and Internet of Things (IoT) increases the requirement for edge computing with low power and relatively high processing speed devices. The Computing-In-Memory(CIM) schemes based on emerging resistive Non-Volatile Memory(NVM) show great potential in reducing the power consumption for AI computing. However, the device inconsistency of the non-volatile memory may significantly degenerate the performance of the neural network. In this paper, we propose a low power Resistive RAM (RRAM) based CIM core to not only achieve high computing efficiency but also greatly enhance the robustness by bit line regulator and bit line weight mapping algorithm. The simulation results show that the power consumption of our proposed 8-bit CIM core is only 3.61mW (256*256). The SFDR and SNDR of the CIM core achieve 59.13 dB and 46.13 dB, respectively. The proposed bit line weight mapping scheme improves the top-1 accuracy by 2.46\% and 3.47\% for AlexNet and VGG16 on ImageNet Large Scale Visual Recognition Competition 2012 (ILSVRC 2012) in 8-bit mode, respectively.
\end{abstract}


%
\IEEEpeerreviewmaketitle

\section{Introduction}
%
%
%
%
\IEEEPARstart
{I}{n} the past decade, with internet of things, cloud computing, computer vision, and artificial intelligence becoming increasingly connected to do perception, cognition, decision, and interaction, sensing devices in intelligent products are going to be the key interfaces to the real world. However, communication, storage, information retrieval, computation, and recognition will face great challenges due to the extremely large amount of sensing data. Because of the separation of the data acquisition, processing, and analysis, the conventional intelligent systems are suffering from problems like high construction cost, high power consumption, low efficiency, and long latency\cite{AIoT}. To address these issues, the majority of AI computations will be moved to light-weight IoT devices\cite{EdgeComputing}. Nevertheless, Moore’s Law has come to the end and the processor performance will be benefited little from Complementary Metal Oxide Semiconductor (CMOS) technology node scaling down. Therefore, we have to design new hardware architectures and software algorithms to meet the requirement of the perception, computation, and storage at the end devices with limited computation capability and storage resources. 


The high density and low power emerging resistive Non-Volatile Memory (NVM)\cite{PCM2,PCM1,STT-MRAM1,STT-MRAM2,STT-MRAM3,STT-MRAM4,RRAM1,RRAM2,RRAM3} which enables massive parallel Computing In-Memory (CIM) is a promising candidate to solve the above-mentioned issues\cite{NS-CIM,nvCIM,CIM_with_NVM}. The majority of works are utilizing the multilevel resistance of the resistive memory for both storage and computation \cite{3,2,1,4}. For example, Hewlett Packard Laboratories (HPL) proposed a Dot Product Engine (DPE) with the inverting amplifier\cite{Dot_Engine}. \cite{ISAAC} designed In-Situ Analog Arithmetic in Crossbars (ISAAC) which utilizes eight 4 level RRAM cells to represent 16-bit weight. Though resistive NVM provides a potential solution as the CIM unit, its non-ideal properties greatly degenerate the reliability of the system. A few widely known properties of resistive NVM are the non-linear resistance value with different biasing voltage, level to level resistance variation, and cell to cell resistance variation, which will cause significant errors in quantization, resulting in the accuracy loss in the network. To reduce the mapping errors and improve the linearity of the CIM system, a more reliable design is needed which may be at the cost of the increasing of the computing energy.\par
\cite{SINWP0},\cite{SINWP1} proposed Serial-Input Non-Weighted Product (SINWP) whose inputs are modulated by time instead of the analog voltage, which will address the non-linearity issue caused by the biasing voltage. However, the digital-to-time converter will greatly increase the computing time at high data width.\cite{Active_integrator_scheme_with_amplifier} proposed a novel Multiple Binary RRAM with Active Integrator (MBRAI) CIM core architecture, where multiple binary RRAM cells are used to store one weight. MBRAI could save a lot of power because binary code is used at the input instead of a time signal. Therefore, it requires only n CIM computations instead of $2^n$. The n-bit input data are sequentially computed by the CIM core and weighted at the output neurons, which greatly improves the linearity because of the identical input voltage. However, the power consumption of this scheme is dominated by the operational amplifier (\textgreater 95\%)\cite{Active_integrator_scheme_with_amplifier}, which reduces the computing efficiency of the CIM core to 0.61 Tera-Multiply-Accumulates per Second per Watt (TMACs/s/W). What's more, the accuracy is still be influenced by the quantization and the inconstancy of the resistive NVM cells. To reduce power consumption, a CIM core with regulated passive integrators is proposed in this paper. A pseudo-binary weight quantization and bit line weight mapping method aimed at solving the resistance inconsistency is also introduced. The simulation results show that the power consumption of the proposed 256*256 CIM core in 8-bit mode is reduced by 98.2\% compared with MBRAI.\par
The rest of this paper is organized in the following manner. Section II introduces the background of CIM with resistive NVMs. Section III shows the proposed circuit, problems brought by resistance inconsistency, and corresponding optimization. Finally, simulation results are presented in Section IV with the conclusion in Section V.


 
\section{BACKGROUND AND RELATED WORKS}

\begin{figure}[h]
	\centering
	\includegraphics[width=50mm]{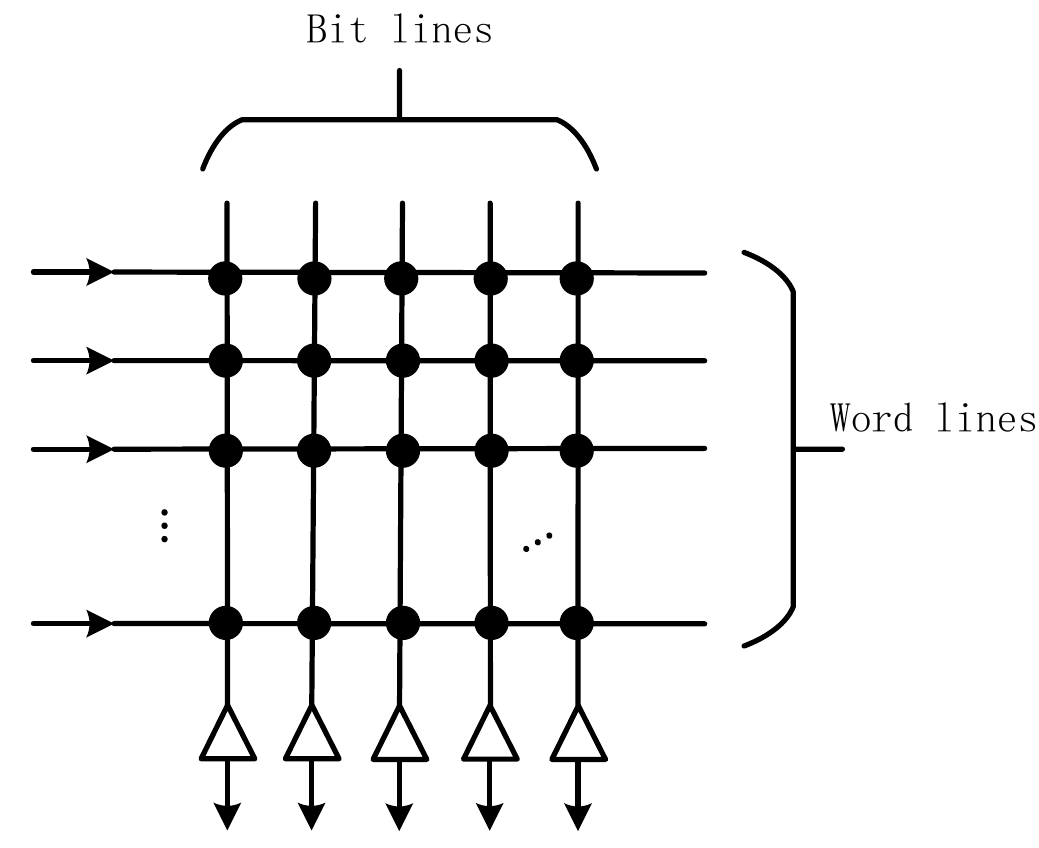}
	\caption{ Microarchitecture of a CIM core. The word lines get input data from the former neurons while the bit lines is to accumulate data at the neurons (triangles). The black dots are the computing unit.}	\label{Microarchitecture of a neuromorphic core}
\end{figure}

The majority of the computations in the neural network are matrix multiplication and accumulate operations, which can be well implemented by crossbar architecture as shown in Fig. \ref{Microarchitecture of a neuromorphic core}. The processing units shown as black dots multiply the input from word lines by the stored weight. The neuron represented by the triangle accumulates the multiplication results at the same bit line.\par
\begin{figure}
	\centering
	\includegraphics[width=7cm]{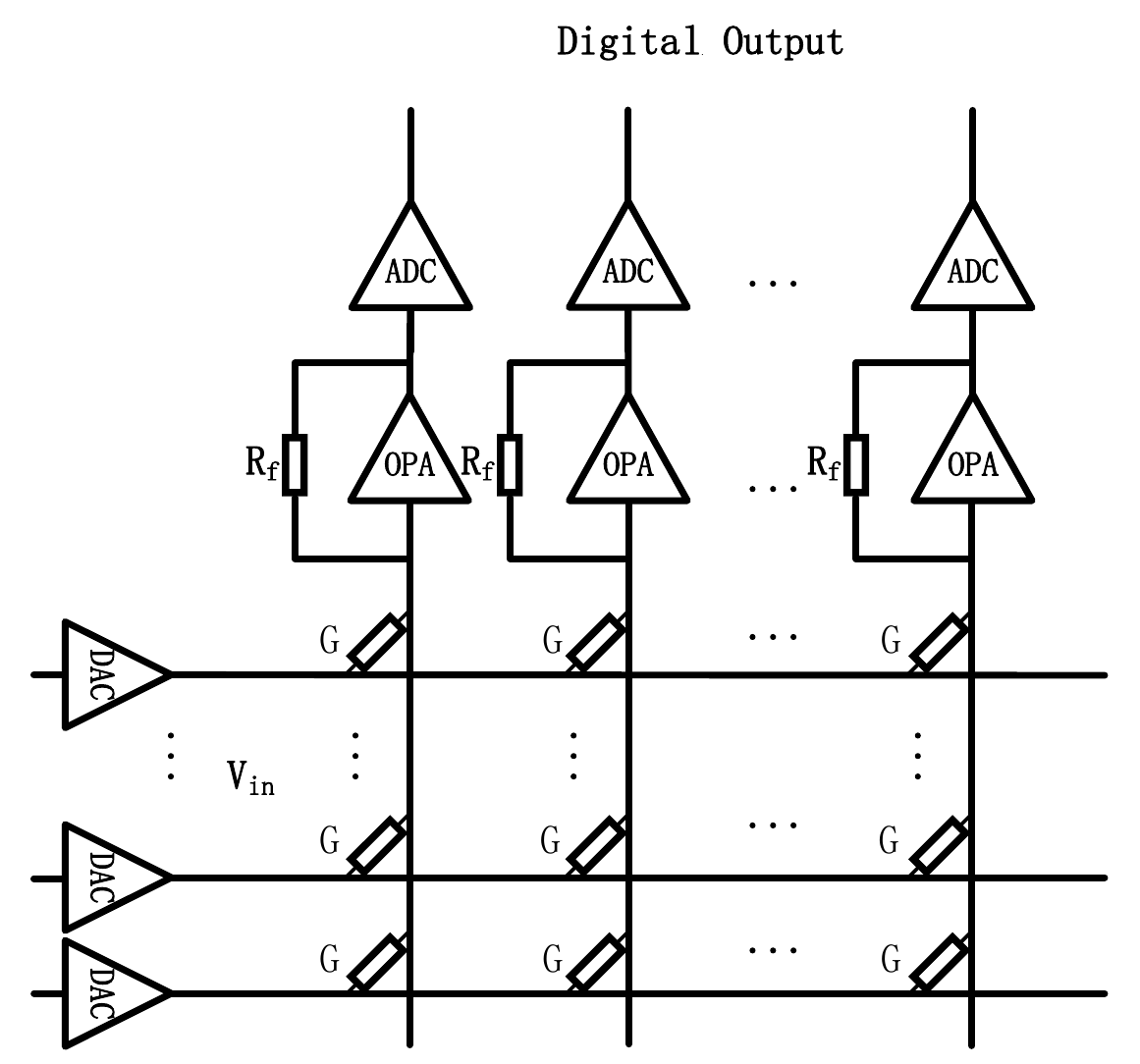}
	\caption{Basic architecture of HPL's Dot-Product Engine}
	\label{Dot product engine}
\end{figure}
\begin{figure}[htbp]
	\centering
	\includegraphics[width=80mm]{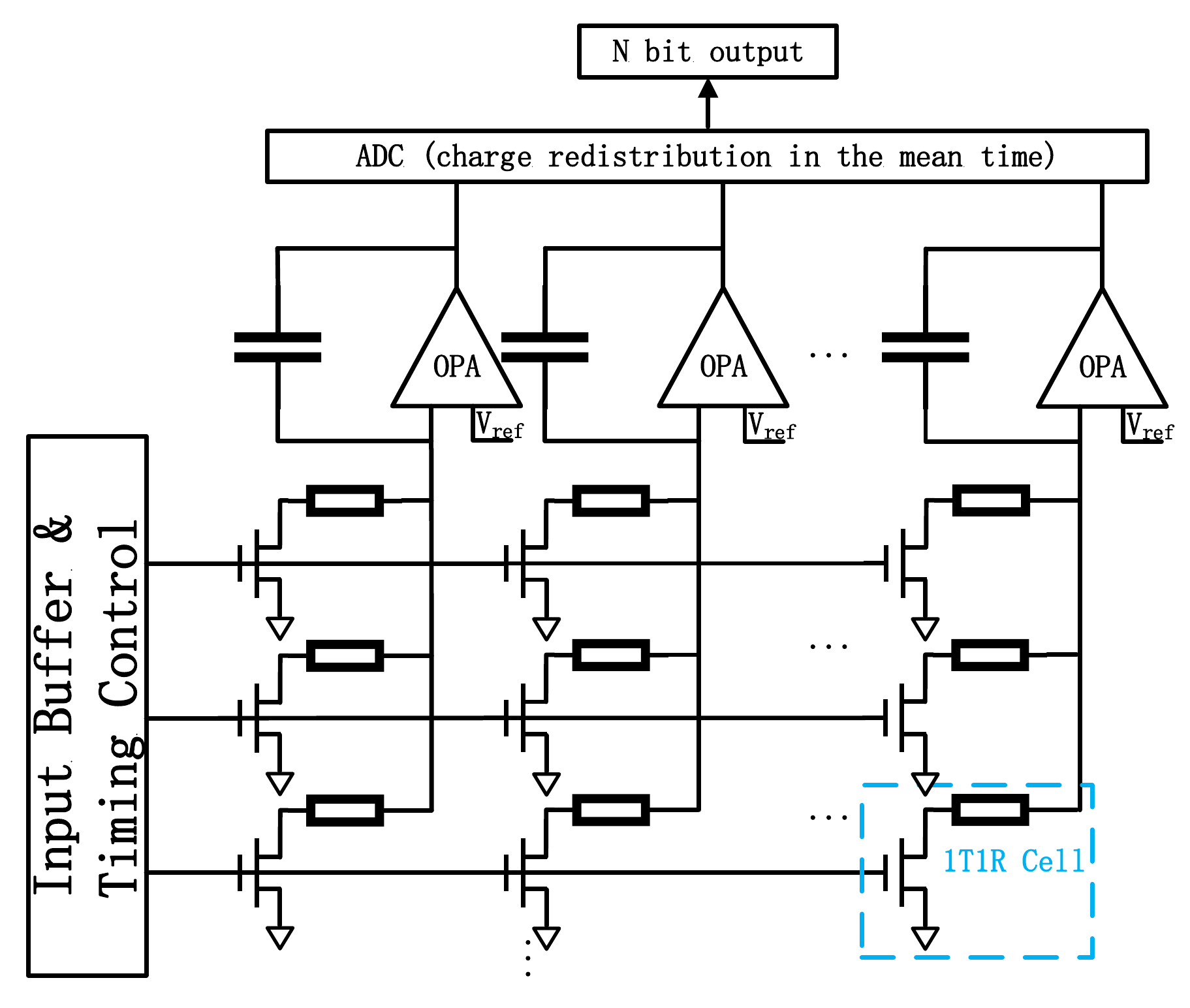}
	\caption{Simplified integration circuit with OPA.}
	\label{Simplified integration circuit with OPA}
\end{figure}

In conventional schemes, the weight of the neural network is stored in SRAM. For example, \cite{conv-RRAM} proposed a 7-bit input and 1-bit weight MAC using a 10T SRAM cell. However, the precision is limited by the 1-bit weight and the area is large due to the SRAM array.
Emerging NVMs which have high density and simpler structure as memory unit will greatly improve the precision of the weight and reduce the core area for its application on CIM. A design of CIM core with NVM in DPE is shown in Fig. \ref{Dot product engine}. It employs memristor crossbar for matrix multiplication where memristor stores the weight by its resistance. Once the input is converted to analog voltage by DAC, the output voltage is determined by the conductance of the resistance as $V_{out} = \sum{V_{in}GR_f}$, where $R_f$ is the feedback resistance, and $G$ is the conductance of the cross-point memristor device. After that, the output voltage is digitalized by the ADC for data transmission. DAC and ADC, which are power-hungry components, are necessary to resist the noise and signal deformation in data transmission. MBRAI proposed in \cite{Active_integrator_scheme_with_amplifier} moves the input DACs to the output and shares the ADC for lower power consumption. RRAM is chosen as the storage unit for its reconfigurability, high density, and low power consumption. However, it is a great challenge to precisely control the resistance value of RRAM. Therefore, MBRAI utilizes n RRAM cells with binary resistance states whose high resistance state (HRS) is 0 and low resistance state(LRS) is 1, to represent an n-bits weight to achieve a high Effective Number of Bits (ENOB) for weight mapping. Fig. \ref{Simplified integration circuit with OPA} is the simplified integration circuit of MBRAI. The input is sent in bit by bit from Least Significant Bit (LSB) to Most Significant Bit (MSB), which is more computing reliable since every bit is identical in computation. The importance of each bit of the input data and network weights are weighted by the charge redistribution at the neurons.\par
Though MBRAI achieves better computing reliability, there are still two critical issues that need to be addressed. Firstly, the power consumed by amplifiers in active integrators accounts for more than 95\% of the energy cost of the whole CIM core. Secondly, the resistance of the RRAM cells has a wide distribution, resulting in significant quantization errors when mapping weights of the neural network into the RRAM array. To address the first issue, a passive integrator is proposed. A regulator is designed to improve the linearity of the integration results. The details will be introduced in Section III.A. To address the second issue, a pseudo-binary quantization and bit line weight mapping method is proposed to reduce the impact of the resistance inconsistency. The details will be introduced in Section III.B.\par

\begin{figure}[htbp]
	\centering
	\includegraphics[height=40mm]{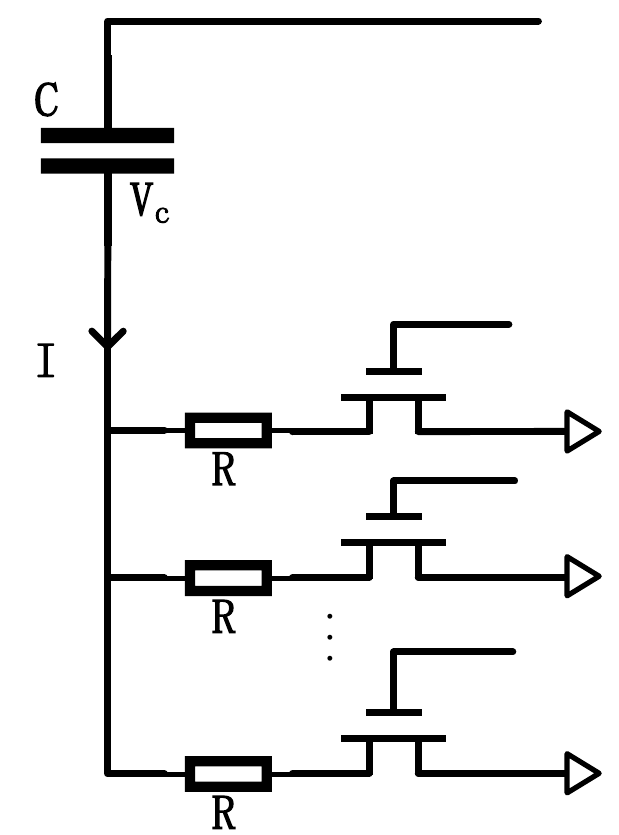}
	\includegraphics[height=40mm]{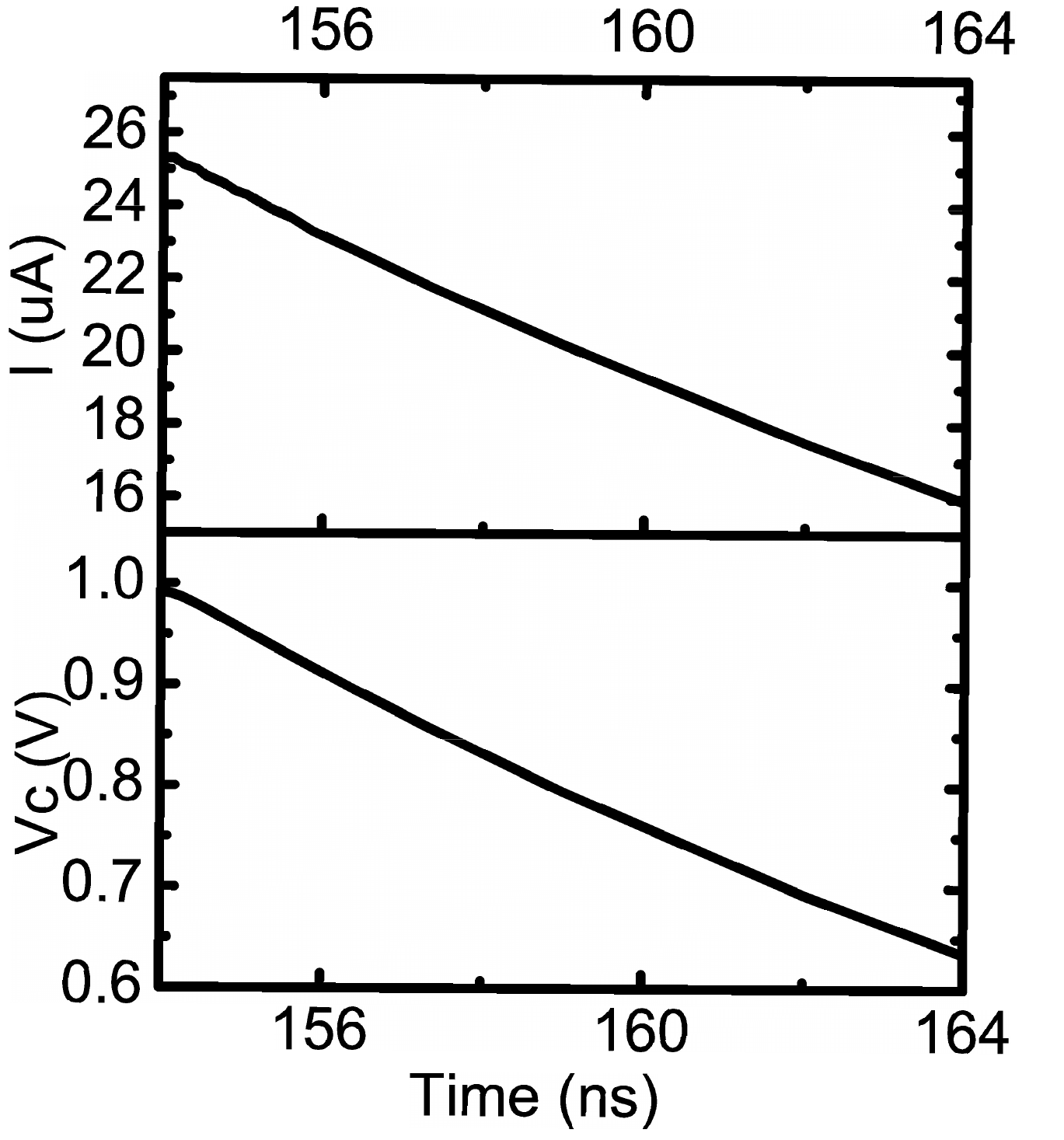}
	\caption{Passive integrator with amplifier removed and its integration process.}
	\label{1R1T_introduction}
\end{figure}
\begin{figure*}[htbp]
	\centering
	\begin{minipage}[t]{1\linewidth}
		\centering
		\includegraphics[width=180mm,height=100mm]{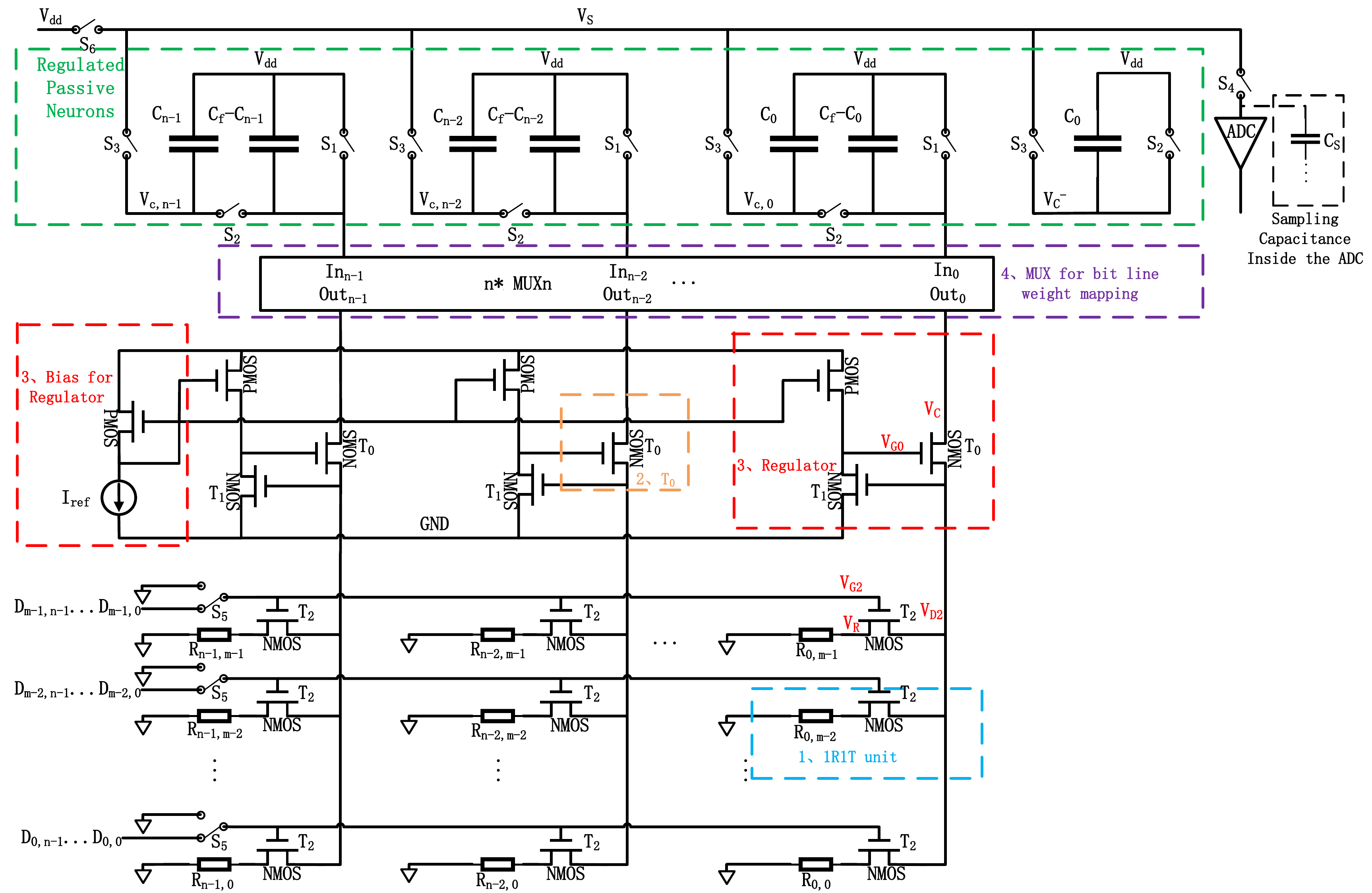}
	\end{minipage}%
	\caption{Integration circuit without amiplifier} 
	\label{Integration circuit}
\end{figure*}
\section{Proposed CIM Core and Mapping Method}

Although a passive integrator can significantly reduce power consumption by removing the amplifier, it has a serious non-linear problem. Fig. \ref{1R1T_introduction} shows the passive integrator circuit and its integration process where the current decreases with the decreasing of the integrating voltage $V_C$. To improve the linearity of the circuit, we design an optimized n-bit integral multiplier shown in Fig. \ref{Integration circuit}:
\begin{enumerate}[]
	\item We switch the position of RRAM and transistor in 1T1R so that the reading voltage on the RRAM cell is mainly determined by the gate and threshold voltages of the transistor. To differentiate from the conventional structure, the new structure is named as 1R1T. 
	\item The saturation current of the transistor in 1R1T can be influenced by the change of the integrating voltage because of the channel length modulation effect. To minimize the impact of the integrating voltage, we add NMOS $T_0$ at bit line to isolate the integrating voltage and drain voltage of 1R1T and thus reduce the variation of the integrating current.
	\item Because the load of the bit line is affected by the number of input lines and the weights' values, the linearity of the circuit is still influenced by the change of the source voltage of T0 (the drain voltage of the 1R1T). Therefore, a regulator is added at T0 to make sure the stability of the drain voltage of 1R1T.
	\item Besides the nonlinearity in the bit line voltage, the cell to cell variation makes the devices' integrating current inconsistent which decreases the robustness of the system. To improve reliability, we propose a pseudo-binary quantization and bit line weight mapping method with corresponding circuit which utilizes the uncertainty of resistive NVM to reduce quantization error.
\end{enumerate}   
\subsection{CIM Core with Regulated Passive Integrator}
\subsubsection{Core Design}
Assuming the n-bit input sequence is $X_1,X_2,...,X_l$ and the weight is $W_1,W_2,...,W_l$, the multiplication and accumulation(MAC) can be expressed
as\\
\begin{equation}
\label{bit by bit computation}
\begin{aligned}
Y&=\sum_{i=1}^{l} X_i W_i = \sum_{i=1}^{l} \sum_{j=0}^{n-1} 2^{j} x_{i,j} W_i \\
&= \sum_{i=1}^{l} \sum_{j=0}^{n-1} 2^{j} \sum_{k=0}^{n-1} 2^{k} x_{i,j} w_{i,k}
\end{aligned}
\end{equation}
where $x_{i,n-1}x_{i,n-2}...x_{i,0}$ and $w_{i,n-1}w_{i,n-2}...w_{i,0}$ is the binary format of $X_i$ and $W_i$ ($x_{i,j},w_{i,k} \in \left(0,1\right) $), respectively. It can be observed from Eq. \ref{bit by bit computation} that there are three consecutive accumulations. The proposed CIM core utilizes n integrator cells to get $\sum_{i=1}^l x_{i,j}w_{i,k}$ by charge integration, and the results are stored in the passive regulated neuron composed by the capacitance array in Fig. \ref{Integration circuit} for charge redistribution to get the $\sum_{i=1}^l x_{i,j}W_{i}$. The $\sum_{i=1}^l x_{i,j}W_{i}$ is also added up by charge redistribution to get $\sum_{i=1}^lX_iW_i$. The integration for the resistances in the same bit line will be finished simultaneously by sharing the integrator so that the MAC can be finished parallelly to achieve a smaller core area and faster computing speed. Multiple neurons are enabled at a time in the integration phase when the inputs are divided into n cycles and calculated from LSB to MSB. After integration and charge redistribution, the data conversion phase is started for neurons to convert the analog results into digital output.\par

\subsubsection{Integral Multiplier}
The word line inputs shown in Fig. \ref{Integration circuit} are sent in once a bit from LSB to MSB. The process of multiplication in the integral multiplier includes the integration phase and the charge redistribution phase. When in the integration phase, $S_2$ is closed, $S_1$, $S_3$, and $S_4$ are open. After the integration, the charge is redistributed with $S_1$ and $S_2$ open and $S_3$ and $S_4$ closed in the charge redistribution phase. Taking $C_{n-1}$ as an example, the integrating voltage after the integration phase is\\
\begin{equation}
\label{integration of one cell}
V_{c,n-1} = V_{c,n-1}^--\frac{V_{D2} \sum_{i=0}^{l-1} D_{i} T}{C_{f}R_{i}}
\end{equation}
where $V_{c,n-1}^-$ is the initial voltage of $C_{n-1}$, $D_i$ is one input bit of the $i_{th}$ input line, $l$ is the number of input lines, T is the integration time, $R_i$ is the equivalent resistance of the 1R1T unit of the $i_{th}$ input line and $V_{D2}$ is the drain voltage of 1R1T unit. 
The capacitances satisfy the following constraint
\begin{equation}
\label{relationship of capacitance}
C_{f}=2C_{n-1}=2^{2}C_{n-2}=2^{3} C_{n-3}=\ldots=2^{n} C_{0}
\end{equation}
Assuming there is only one input line and the initial integrating voltage is set to $V{c}^{-}$, the integrating voltage Vs after one step of charge redistribution is
\begin{equation}
\begin{aligned}
\label{charge redistribution in one bit}
V_{S} &=\frac{V_{c, n-1} C_{n-1}+V_{c, n-2} C_{n-2}+\ldots+V_{c, 0} C_{0}+V_{C}^- C_{0}}{C_{n-1}+C_{n-2}+\ldots+2 C_{0}} \\
&=V_C^{-}-\frac{V_{D2} T}{C_{f}}\left(2^{-1} \sum_{i=0}^{l-1} \frac {D_{i, n-1}}{ R_{i, n-1}}+2^{-2} \sum_{i=0}^{l-1} \frac {D_{i, n-2}}{ R_{i, n-2}}\right.\\
&\left.+\ldots+2^{-n} \sum_{i=0}^{l-1}\frac {D_{i,0}}{ R_{i, 0}}\right)
\end{aligned}
\end{equation}
ADC's sampling capacitance $C_S$ which is connected to the integrator's capacitance array is in the meantime used to add up the n partial products. Let $C_S=C_f$, the new $V_{out}$ is
\\
\begin{equation}
V_{out}=2^{-1}\left(V_{S}+V_{out}^{-}\right)
\end{equation}
where $V_{out}^-$ represents the former voltage of the $C_S$. Assuming the initial voltage of the $C_S$ is $V_{init}$, then after n steps of the charge redistribution, the voltage change is 

\begin{align}
\label{charge redistribution}
\triangle V_{out}=&V_{init}-\left(2^{-n}V_{init}+2^{-n} V_{S, 0}+\ldots+2^{-1} V_{S, n-1}\right)\nonumber\\
=&2^{-n}\left[\left(V_{init}-V_{S, 0}\right)+\ldots+2^{n-1}\left(V_{init}- V_{S, n-1}\right)\right]\nonumber\\
=&2^{-n}\sum_{j=0}^{n-1}2^{j}\triangle V_{S,j}        
\end{align}

where $V_{S,n-1}$ is the $(n-1)_{th}$ integrating voltage of $V_S$ and $\triangle V_{S,j}$ is the voltage change of $V_S$ in the $j_{th}$ integration. As long as the $\triangle V_{S,j}$ is designed to represent the result of the $\sum_{i=0}^l x_{i,j} W_i$, Eq. \ref{charge redistribution} gives the result of $\sum_{i=0}^l \sum_{j=0}^{7} 2^{j} x_{i,j}W_i$.

\subsubsection{Regulated Passive Integrator}
By switching the position of 1T1R in Fig. \ref{Simplified integration circuit with OPA} to 1R1T in Fig. \ref{Integration circuit}, we can get the following equations
\begin{equation}
I = \frac{1}{2} K_2 \left( V_{G2}-V_{R} - V_{th2} \right)^{2}
\label{1T1R_T}
\end{equation}
\begin{equation}
I = \frac{V_{R}}{R}
\label{1T1R_R}
\end{equation}
where $K_2$ is the device parameter of $T_2$, $V_{th2}$ is its threshold voltage, R is the resistance of RRAM device, $V_R$ is resistance's read voltage, $I$ is the integrating current passing through the 1R1T unit. According to Eq. \ref{1T1R_T} \ref{1T1R_R}, we can get
\begin{equation}
V_R = V_{G2}-V_{th2}-\frac{\sqrt{2K_2R(V_{G2}-V_{th2})+1}-1}{K_2R}
\end{equation}
The drain voltage of $T_2$ ($V_{D2}$), which is isolated from the integrating voltage by T0, satisfies the following equation
\begin{equation}
\label{T0}
I_b = \frac{1}{2} K_0 \left(V_{G0}-V_{D2}-V_{th0}\right)^{2}
\end{equation}
where $I_b$ is the integrating current of the bit line, $K_0$ is the device parameter of $T_0$.
The proposed regulator circuit shown in Fig. \ref{Integration circuit} stabilizes the $V_{D2}$ of the 1R1T units by applying a negative feedback. $T_1$ works at the saturation region, which satisfies the following equation
\begin{equation}
I_{ref} = \frac{1}{2} K_1\left(V_{D2}-V_{th1}\right)^{2}
\label{Regulator}
\end{equation}
where $K_1$ is the device parameter of $T_1$, $V_{th1}$ is the threshold voltage of $T_1$. According to Eq. \ref{T0} \ref{Regulator}, we get
\begin{equation}
V_{G0} = V_{th1} + \sqrt{\frac{I_{ref}}{K_1}}+V_{th0}+\sqrt{\frac{I_b}{K_{0}}}
\end{equation}
\begin{equation}
V_{D2} = V_{th1} + \sqrt{\frac{I_{ref}}{K_1}}
\end{equation}
Since $I_{ref}$ is a constant, the drain voltage $V_{D2}$ of the 1R1T unit is stabilized by the regulator.


\subsection{Pseudo-binary Quantization and Bit Line Weight Mapping Method}
As the weight of the neural network is quantized to n bits rather than a continuous value, the quantization errors when mapping the weight of the neural network into the CIM system will influence the accuracy of inference. What's more, the resistance distribution of resistive NVM may worsen the quantization. Therefore, it's necessary to discuss the quantization method and the corresponding errors in this section. To reduce the quantization error caused by the cell to cell variation, a pseudo-binary quantization and bit line weight mapping method is proposed.\par
\subsubsection{Quantization Error with NVM}
Quantization is an important method for compressing the neural network and accelerating the computation speed, among which uniform quantization is a basic one. The typical quantizer of uniform quantization can be expressed as\\ 
\begin{equation}
Q(x)=\Delta \cdot\left\lfloor\frac{x}{\Delta}+\frac{1}{2}\right\rfloor
\end{equation}
where $\Delta$ is the quantization step size of some value, x is the value to be quantized. 
When the quantization step size ($\Delta$) is small relative to the variation in the signal being quantized, it is simple to show that the mean squared error which is also called the quantization noise power produced by such a rounding operation will be $\frac{\Delta^{2}}{12}$. The calculation process is\\
\begin{equation}
Q E=\int_{0}^{\frac{\Delta}{2}} \frac{x^{2}}{\frac{\Delta}{2}} d x=\frac{\Delta^{2}}{12}
\end{equation}
The maximum ($w_{max}$) and the minimize ($w_{min}$) of the data range and the quantization bits n determine the quantization step size ∆ since they usually have the relationship\\
\begin{equation}
\Delta \times 2^{n}=\left(w_{\max }-w_{\min }\right)
\end{equation}
Considering the resistance distribution, the practical non-linear quantizer is shown as follows\\
\begin{equation}
\sum_{i=1}^{2^{n}} \Delta_{i}=\left(w_{\max }-w_{\min }\right)
\end{equation}
Assuming the resistance distribution is a general normal distribution represented as\\
\begin{equation}
f\left(x | \mu, \sigma^{2}\right)=\frac{1}{\sqrt{2 \pi \sigma^{2}}} e^{-\frac{(x-\mu)^{2}}{2 \sigma^{2}}}
\end{equation}
The Probability Density Function (PDF) of the quantization error is the noncentral chi-squared distribution with one degree of freedom. Then, the mean value of the quantization error is given by\\
\begin{equation}
\mu=k+\frac{\lambda}{12}=1+\frac{\mu^{2}}{12}=1+\frac{\Delta^{2}}{12}=1+\frac{\left(w_{\max }-w_{\min }\right)^{2}}{12 \times 2^{2 n}}
\end{equation}
and the variance of the quantization error is\\
\begin{equation}
\begin{aligned}
\sigma^{2}&=2(k+2 \lambda)=2+4 \mu^{2}=2+4 \Delta^{2}\\
&=2+\frac{\left(w_{\max }-w_{\min }\right)^{2}}{2^{2(n-1)}}
\end{aligned}
\end{equation}\par
As we can see, the quantization error is greatly increased when there is a distribution in resistance. Since the quantization errors are accumulated in the network, the accuracy will be greatly reduced.
\subsubsection{Resistance Measurement}The proposed quantization and mapping method needs the resistance value of the RRAM array in LRS, so we firstly set all memory units to LRS and read the resistance of the RRAM array by ADC in resistance reading phase. The reading process consists of integration phase and charge redistribution phase when the switches are set different from multiplication. For example, when reading the resistance unit $R_{n-1,m-1}$ in Fig. \ref{Integration circuit}, the switches in the same bit line with $R_{n-1,m-1}$ are used while the others stay open. In the integration phase, $S_1$, $S_3$, and $S_4$ are open, $S_2$ is closed and the input of $(m-1)_{th}$ word line is 1 while the others are 0. The integration result is a typical result of Eq. \ref{integration of one cell} where $l$=1, D=1. When ADC read the integrating voltage during the charge redistribution, $S_2$, $S_3$, and $S_4$ are closed and $S_1$ is open. The voltage read by ADC is 
\begin{equation}
V_{out} = \frac{V_{init}+V_{S}}{2}
\end{equation}
where $V_{init}$ is the initial voltage for both sampling capacitances and integration capacitances, and $V_S$ is the integration result. The integration process satisfies
\begin{equation}
V_{init}-V_S = \frac{IT}{C_f}=\frac{V_{D2}T}{R C_f}
\end{equation}
where I is the integrating current passing through 1R1T unit, T is the integration time, R is 1R1T's resistance and $V_{D2}$ is the read voltage of the bit line. Therefore, we can get 
\begin{equation}
R = \frac{V_RT}{2C_f\left(V_{init}-V_{out}\right)}
\end{equation}
\par
\subsubsection{Quantization and Mapping Method}Since the normalized resistance in LRS is not exactly digital 1, a pseudo-binary code whose importance of bits from MSB to LSB is still the same as the conventional binary code is proposed in our mapping schemes. The main difference is that the value of the pseudo-binary code is related to the resistance of the memory unit, which is given by\\
\begin{equation}
\hat{w}=r_{n-1} \times 2^{n-1}+r_{n-2} \times 2^{n-2}+\ldots+r_{0} \times 2^{0}
\end{equation}
where for the LRS, $r_i$ is the normalized resistance of the $i_{th}$ bit (mean value is 1) of the weight. For the high resistance, since the resistance can be much larger than the low resistance, $r_i$ is set as 0, and the uncertainty of the high resistance is ignored in this paper. The weight quantization procedure is from MSB to LSB and we define the condition as follows\\
\begin{equation}
\begin{aligned}
q =& !\left[ {\left(r_{i} \times 2^{i-1}-w_{r e s}>0.5\right)} \right. \\
&\left. {\left|\left(r_{i}<=0.5\right)\right|\left(r_{i} \times 2^{i-1}>2 \times w_{r e s}\right)} \right] \\
\end{aligned}
\label{quantization equ}
\end{equation}
where $r_i$ is the $i_{th}$ bit of normalized memory resistance, and $w_{res}$ is the remaining weight after partial quantization. The component $\left(r_{i}<=0.5\right)$ is to abandon the device with too large resistance in LRS and $r_{i} \times m_{i}-w_{res}>0.5$ is to check if the remaining weight is larger than the product of the importance of the bit and the resistance of the memory. The component $r_{i} \times m>2 \times w_{res}$ is to minimize the quantization error of the LSB. Because of the memory resistance distribution, $\left|r_{0} \times m_{0}-w_{res}\right|$ could be larger than $w_{res}$. In other words, the memory should be in high resistance in case $\left|r_{0}-w_{res}\right|>w_{res}$ to minimize the quantization errors.\par
However, the initial memory sequence may not be the best solution to minimize the quantization error. For example, assuming w=13.4, and four memory units with normalized resistance 1.05, 1.1, 1.125, 0.93 are used to quantize the weight. The conventional binary code may give a quantization error of 0.4 (4'b1101). Based on the given sequence, the resistance states of the four cells are low, low, high, low, which reduce the quantization error to -0.33. Furthermore, if we switch the third cell with MSB, the resistance of the four cells will be 1.125, 1.1, 1.05, and 0.93. In such a sequence, the resistance states of the four cells can be set to low, low, high, and high to minimize the quantization error to 0. This example shows that the sequence of the memory units is very important to minimize the quantization error.\par
The traversal algorithm can be used to search all possible sequences, and the sequence for the minimal quantization error is picked and configured in the chip. However, it may require a long searching time and its computation complexity is O($n!$). Moreover, the cells in the same bit line should be in the same sequential position. Assuming the size of the weight matrix need to be quantized is R*1 and the size of RRAM array is R*C where C means the weight is quantized to C bits, we propose a greedy mapping algorithm whose loss when quantizing the $i_{th}$ bit is defined as
\begin{equation}
\begin{aligned}
\label{loss}
\text {loss}=\max\limits_{j \in R} \left(w_{i-1,j}-\hat{w}_{i,j}\right) \sum\limits_{j=1}^{j=R}\left(w_{i-1,j}-\hat{w}_{i,j}\right)^{2}
\end{aligned}
\end{equation}
where $w_{i-1,j}$ is the remaining value of the $j_{th}$ weight after partial quantization and $\hat{w}_{i,j}$ is the value quantized by the pseudo-binary quantization method in the $i_{th}$ bit of the $j_{th}$ row. Eq. \ref{loss} has taken both the worst case and the average case of the searching results into consideration. This bit line selection is started from MSB which influences the mapping result most to LSB. The algorithm traverses the remaining bit lines and chooses the bit line with minimum loss as the $i_{th}$ bit. To apply the algorithm in the circuit, the n* MUXn in Fig. \ref{Integration circuit} is used for switching the connection between the bit lines of the RRAM array and the integrators. When mapping the weights to the core, all the RRAMs are set to LRS at first. Then, according to the proposed mapping method, RRAMs with value 0 are set to HRS. By using this bit line weight mapping method, the computation complexity is reduced from O($n!$) to O($n^2$).\par 

\begin{figure}[h]
	\subfigure[]{\label{intergration}
		\begin{minipage}[t]{1\linewidth}
			\centering
			\includegraphics[height=40mm,width=202pt]{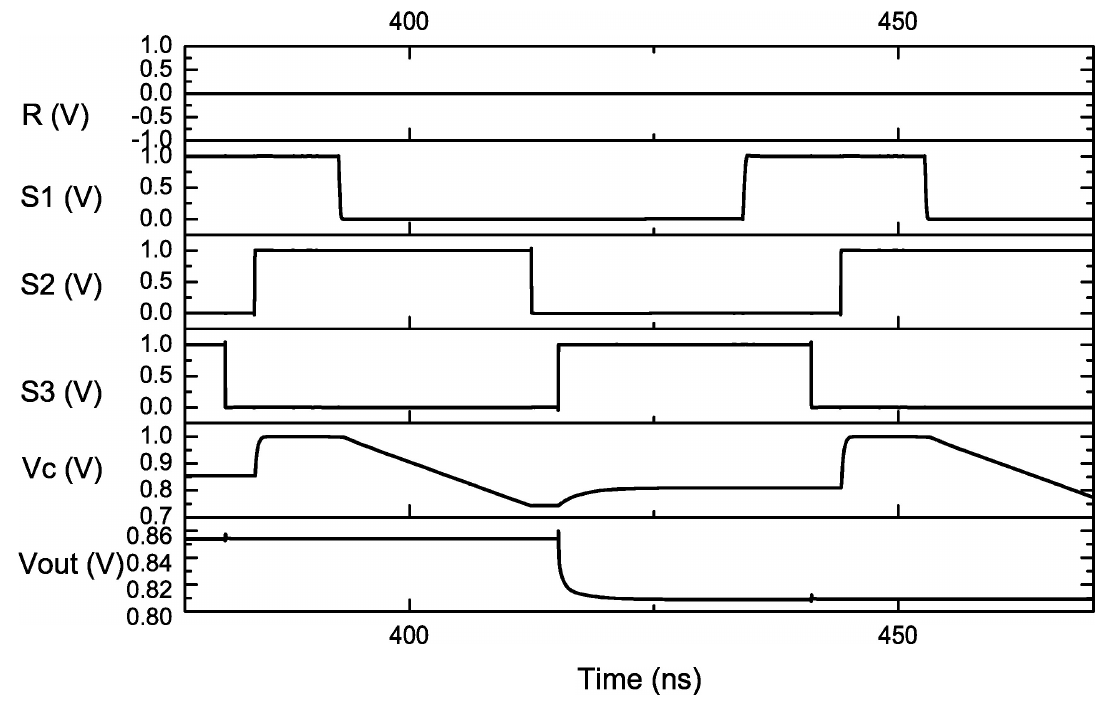}
		\end{minipage}
	}
	\subfigure[]{\label{function verification}
		\begin{minipage}[t]{1\linewidth}
			\centering
			\includegraphics[height=40mm,width=200pt]{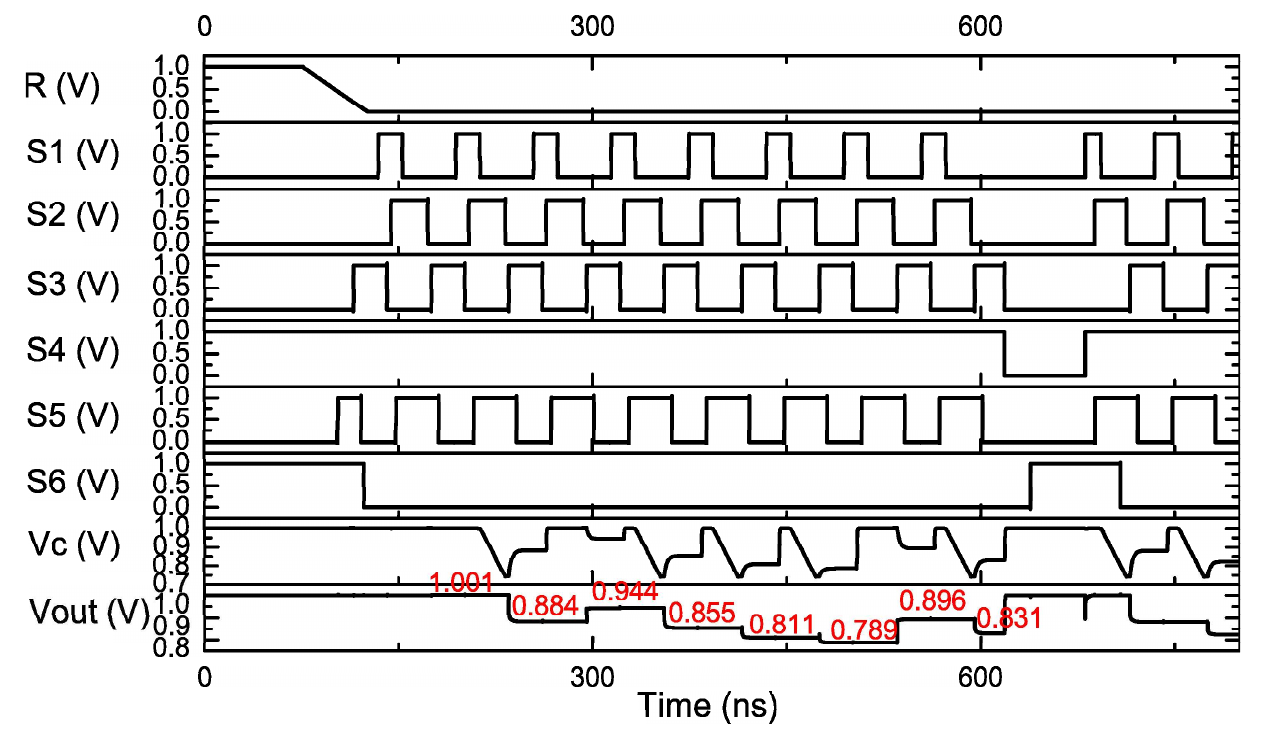}
		\end{minipage}
	}
	
	\caption{Transient simulation results of (a) integration phase and charge redistribution phase for one bit of input (b) the core's multiplication process of 8'b10111010 as input and 8'b11101100 as weight. }
	\label{functional verification}
\end{figure}
\begin{figure}[h]
	\centering
	\includegraphics[height=45mm,width=200pt]{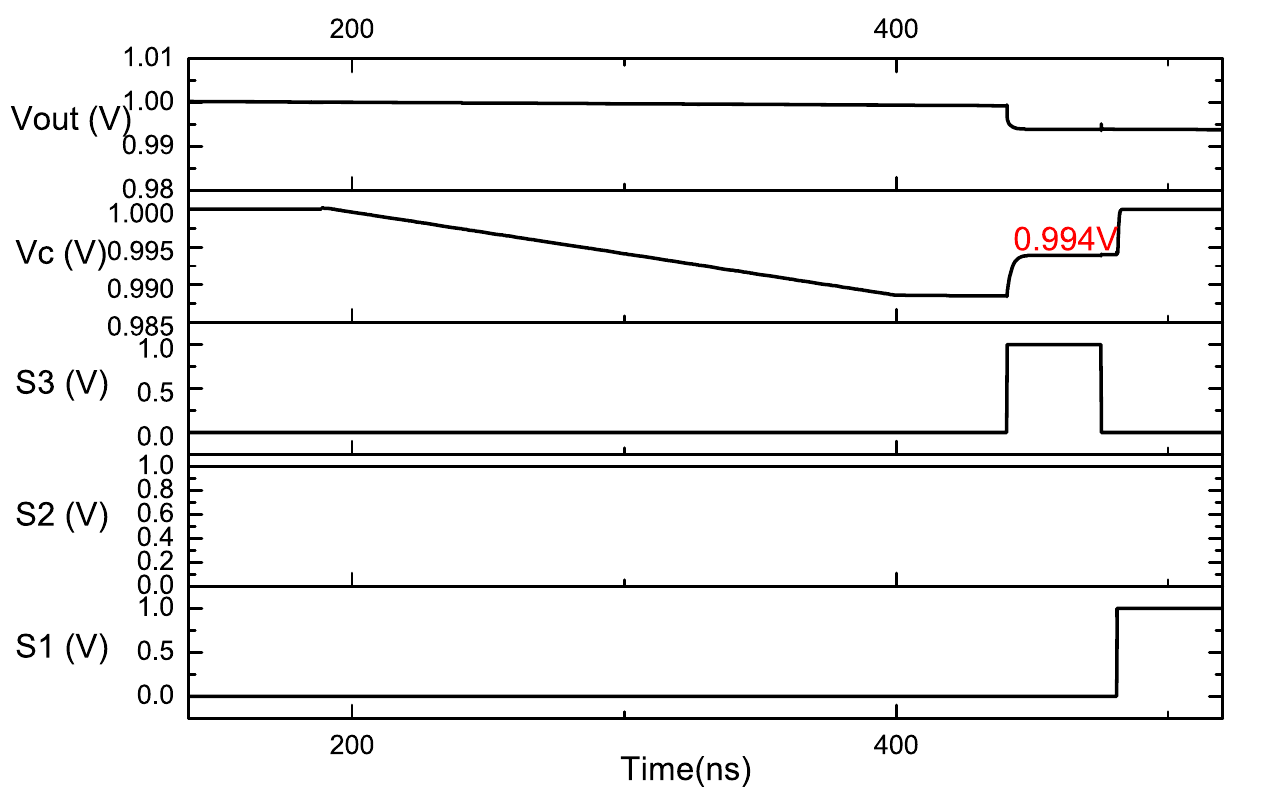}
	\caption{Process of resistance reading where the state of the switches controlling it is presented}
	\label{resistance measurement}
\end{figure}

\section{Simulation Results}
In this section, we do the functional verification of the multiplication and resistance reading process. Then we evaluate the circuit with dynamic performance, energy cost on circuit level and compare it with other CIM schemes on core level and network level. Finally, we present the robustness of the circuit. The circuit simulations are done in Cadence Analog Mixed Signal (AMS) with a 45nm generic Process Design Kit (PDK) and the network simulations are done on caffe platform.
\subsection{Functional Verification}
\subsubsection{Multiplication Process Verification}
We simulate the computing process shown in Fig. \ref{functional verification} to check the correctness of the proposed circuit in 8-bit mode. Fig. \ref{intergration} presents the integration phase in an integrator, the integrating voltage $V_C$ shown in Fig. \ref{Integration circuit} is reset to 1V at 384 ns, and the integration phase starts at 393 ns. After 20 ns, the integration phase is completed and $V_C$ is decreased to 745.2mV. Then the charge redistribution starts at 415 ns. When charge redistribution is done, the 8 integrating voltages are converted to $V_{out}$. After that, $V_C$ is reset to 1V for the next integration. Fig. \ref{function verification} shows the whole multiplication process of an 8-bit input (8'b10111010) and 8-bit weight (8'b11101100). The input 
\begin{table}[t]
	\centering
	\caption{CIM core performance comparison between MBRAI and the proposed}
	\resizebox{50mm}{10mm}{
		\begin{tabular}{|c|c|c|c|c|}
			\hline
			&MBRAI\cite{Active_integrator_scheme_with_amplifier} &Proposed\\ \hline
			Supply Voltage & 1.1V & 1V \\ \hline
			Computing Speed & 1.85M/s &  1.85M/s \\ \hline
			SFDR & 67.42dB & 59.13dB  \\ \hline
			SNDR & 45.48dB &  46.13dB \\ \hline
			ENOB & 7.26bit &  7.37bit  \\ \hline
			
	\end{tabular}}
	\label{Performance comparison}
\end{table}

\begin{table}[]
	\centering
	\caption{Energy cost comparison between MBRAI and proposed CIM core}
	\resizebox{80mm}{12mm}{
		\begin{tabular}{|c|c|c|c|c|c|c|}
			\hline
			\multicolumn{2}{|c|}{  MBRAI\cite{Active_integrator_scheme_with_amplifier}} & \multicolumn{2}{|c|}{Proposed}  \\ \hline
			Technology & 45nm & Technology &  45nm \\ \hline
			Supply Voltage & 1.1V & Supply Voltage &  1V  \\ \hline
			System Clock & 16.7MHz & System Clock & 16.7MHz  \\ \hline
			Integral Amlifier & 0.22mW & Regulator circuit & 1.11uW\\ \hline
			Core(256*256) & 199.68mW & Core(256*256) & 3.61mW  \\ \hline
	\end{tabular}}
	\label{Energy cost comparison}
\end{table}
\begin{table*}[h]
	\caption{Core level comparison between the proposed scheme and other CIM schemes}
	\resizebox{\textwidth}{20mm}{
		\begin{tabular}{|c|c|c|c|c|c|c|c|}
			\hline
			Structure & Technology & Crossbar Size & Weight/Data Bit & Throughout(GMACS) & Power(mW) & Efficiency(TMACs/s/W)    \\ \hline
			\multirow{2}*{SINWP\cite{SINWP0}\cite{SINWP1}} & \multirow{2}*{55nm} & \multirow{2}*{256*512} & fixed-3/fixed-1 & --- & --- &  53.17    \\ \cline{4-7}
			&  &  & fixed-3/fixed-2 & --- & --- &   21.9   \\  \hline
			\multirow{3}*{MBRAI\cite{Active_integrator_scheme_with_amplifier}} & \multirow{3}*{45nm} & \multirow{4}*{256*256}& fixed-3/fixed-1 & 1524 & 19.6 &  77.76   \\ \cline{4-7}
			&  &  & fixed-3/fixed-2 & 1040 & 26.8 &   38.8   \\ \cline{4-7}
			&  &  & fixed-8/fixed-8 & 121.4 & 199.68 &   0.61   \\ \hline
			\multirow{3}*{A 22nm 2Mb ReRAM CIM Macro\cite{22nm_2Mb_RRAM_CIM}}& \multirow{3}*{22nm} & \multirow{3}*{512*512} & fixed-2/fixed-1 &--- & --- &  121.38    \\ \cline{4-7}
			&  &  & fixed-4/fixed-2 &---&---&45.52    \\ \cline{4-7}
			&  &  & fixed-4/fixed-4 &---&---&28.93   \\ \hline
			\multirow{3}*{Proposed} & \multirow{3}*{45nm} & \multirow{4}*{256*256} & fixed-2/fixed-2 & 1092.2 & 1.975 &   553.01 \\ \cline{4-7}
			&  &  & fixed-4/fixed-4 & 546.1 & 2.66 &   205.30\\ \cline{4-7}
			&  &  & fixed-8/fixed-8 & 121.4 & 3.61 &    33.63  \\ \hline
			A CIM SRAM Macro in 7nm FinFET CMOS\cite{351}&7nm&4kb&fixed-4/fixed-4&186.2&1.06&175.5\\ \hline
	\end{tabular}}
	\label{Core level comparison}
\end{table*}
\begin{table*}[]
	\centering
	\caption{Accuracy and energy estimation of different RRAM-based scheme}
	\resizebox{\textwidth}{12mm}{
		\begin{tabular}{|c|c|c|c|c|c|c|c|c|}
			\hline
			Network & The Number of Operations & Structure & System Frequency & Data Bit & Crossbar Size & top-1 error Rate & Energy(uJ/img) & Saving(\%) \\ \hline
			\multirow{3}* {LeNet on MNIST} & \multirow{3}* {0.42M} & BNN+ADCs \cite{BNN_ADCs} & 100MHz & 1 & 128*128 & 1.40\% & 6.68 & 99.81\% \\ \cline{3-9}
			&  & MBRAI\cite{Active_integrator_scheme_with_amplifier} & 25MHz & 8 &  256*256 & 0.97\% & 0.71 &  98.17\% \\ \cline{3-9}
			&  & Proposed & 16.7MHz & 8 & 256*256 & 0.90\% & 0.013 & ---\\ \hline
			\multirow{3}* {AlexNet on ILSVRC 2012} & \multirow{3}* {720M} & BNN+ADCs \cite{BNN_ADCs} & 100MHz & 1 & 128*128 & 73.90\% & 5.42E+03 & 99.69\% \\ \cline{3-9}
			&  & MBRAI\cite{Active_integrator_scheme_with_amplifier} & 25MHz & 8 &  256*256 & 44.16\% & 1.23E+03 & 98.64\% \\ \cline{3-9}
			&  & Proposed & 16.7MHz & 8 & 256*256 & 43.60\% & 16.65 &  ---\\ \hline
	\end{tabular}}
	\label{Accuracy and energy estimation of different RRAM scheme}
\end{table*}
sequence is sent in from LSB to MSB and after 8 cycles of integration and charge redistribution, the output voltage $V_{out}$ is 831.6mV. Then ADC converts it to digital result as 8'b10101011. The theoretical results of the output voltage and digital result are 831.5mV and 8'b10101011, respectively. Therefore, the design achieves its functional requirement.

\subsubsection{Resistance Measurement Verification}
Fig.\ref{resistance measurement} presents the resistance measuring process of one 1R1T unit where the state of the switches in the same bit line is simulated. The output voltage is set to 1V at first and the integration phase is started at 190 ns. Since only one 1R1T is working, the integrating current is small and thus the integrating time is set to 110 ns which is much longer than that of MAC operation. After integration, the sampling phase (i.e. the charge redistribution phase) starts at 440ns and the output voltage is 0.994 V. Then the ADC converts it to digital output.

\subsection{Performance Evaluation}
\subsubsection{Circuit Level Performance}
Table \ref{Performance comparison} shows the dynamic performance comparison between MBRAI and the proposed core. The computing speed, SFDR, SNDR, ENOB of the proposed CIM core are 1.85M/s, 59.13dB, 46.13dB, and 7.37bit, which are close to the performance indicators of MBRAI. Table \ref{Energy cost comparison} gives the power cost comparison between the proposed scheme and MBRAI. MBRAI consumes 0.22 mW on amplifiers for stable read voltage while the proposed circuit only consumes 1.11uW on the regulator circuit, and the total power consumption of the core(256*256) is reduced by 98.2\%. 

\subsubsection{Core Level Comparison}
The core level comparison between the proposed scheme and the other CIM core schemes is shown in Table \ref{Core level comparison}. The simulation results show that the proposed design achieves energy efficiency as high as 553.01 TMACs/s/W in 2-bit input 2-bit weight pattern, 205.30 TMACs/s/W in 4-bit input 4-bit weight pattern, and 33.63 TMACs/s/W in 8-bit input 8-bit weight pattern. Compared with MBRAI, whose energy efficiency is 77.76 TMACs/s/W in 1-bit input 3-bit weight pattern, 38.8 TMACs/s/W in 2-bit input 3-bit weight pattern, and 0.61 TMACs/s/W in 8-bit input 8-bit weight pattern, the proposed scheme achieves much higher energy efficiency (55.13 times in 8-bit input 8-bit weight pattern). Though \cite{351} achieves a low average power consumption in fixed-4 input and fixed-4 weight pattern, the throughout of the core is limited by the rate coding scheme. Meanwhile, the power consumption in \cite{351} will increase with the input value increase, which may achieve a much higher power consumption in practice. Comparing with other CIM schemes, the proposed CIM core achieves better energy efficiency.

\subsubsection{Network Level Comparison}
The accuracy and energy estimation comparison between the proposed scheme and other RRAM based schemes is shown in Table \ref{Accuracy and energy estimation of different RRAM scheme}. Though the binary CIM scheme performs well on small-scale networks, the performance of this scheme on large-scale networks is much worse than the multibit based schemes because of its 1-bit quantization. When considering the energy cost, our scheme reduces 99.81\% of inference energy per image compared with the binary CIM scheme and 98.17\% compared with MBRAI for LeNet on MNIST. The proposed scheme also reduces 99.69\% inference energy per image compared with the binary CIM scheme and 98.64\% compared with MBRAI for AlexNet on ILSVRC 2012. Therefore, by abandoning the amplifiers, the proposed scheme achieves much lower inference energy cost.\par

\subsection{Robustness Analysis}
\subsubsection{Linearity Analysis}
\begin{figure}[]
	\subfigure[]{\label{1R1T}
		\begin{minipage}[]{0.3\linewidth}
			\centering
			\includegraphics[width=28mm,height=25mm]{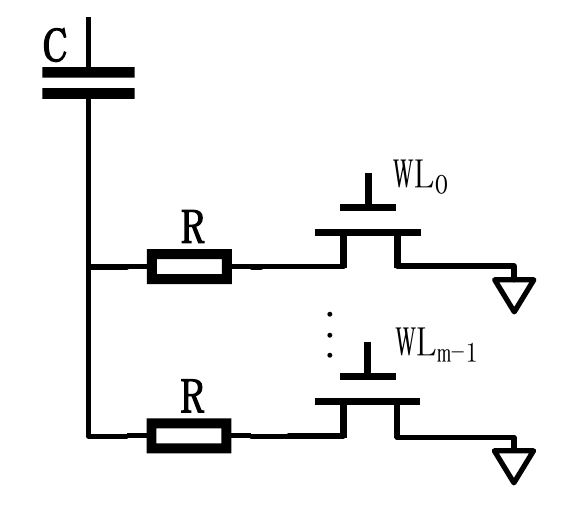}
		\end{minipage}
		\begin{minipage}[]{0.3\linewidth}
			\centering
			\includegraphics[width=28mm,height=25mm]{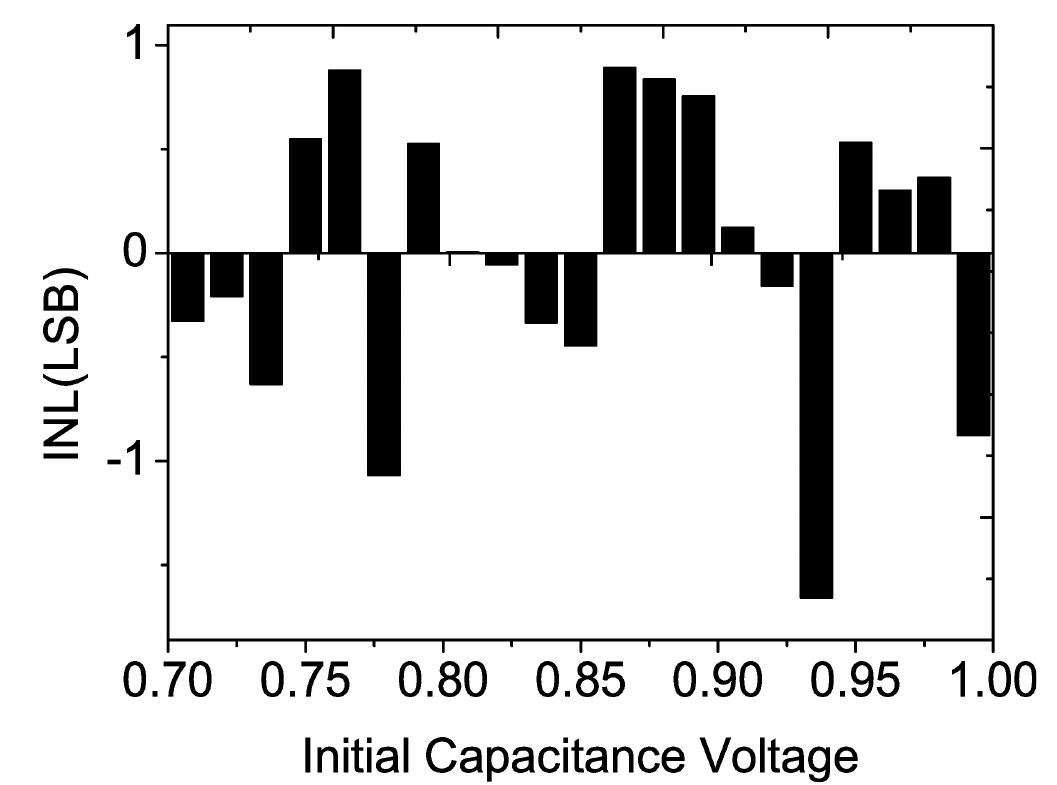}
		\end{minipage}
		\begin{minipage}[]{0.3\linewidth}
			\centering
			\includegraphics[width=28mm,height=25mm]{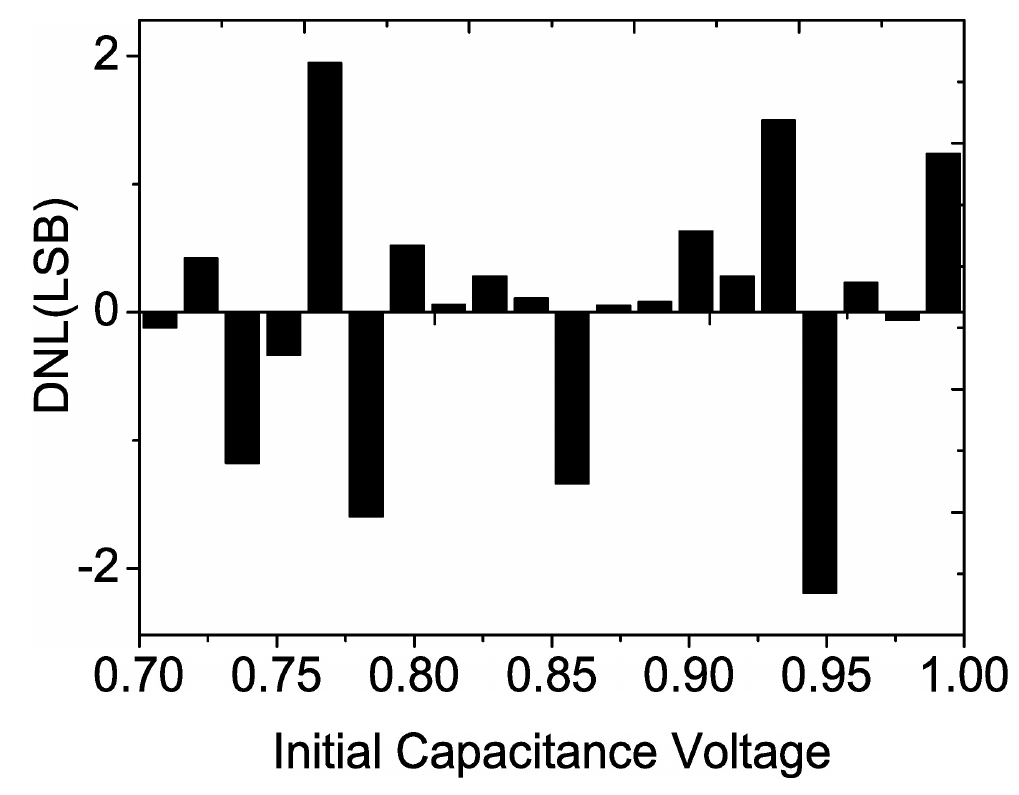}
		\end{minipage}
	}

	\subfigure[]{\label{no_T0}
		\begin{minipage}[]{0.3\linewidth}
			\centering
			\includegraphics[width=28mm,height=25mm]{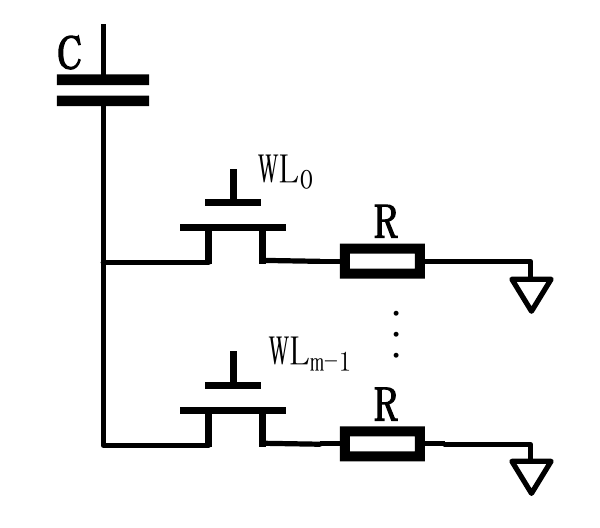}
		\end{minipage}
		\begin{minipage}[]{0.3\linewidth}
			\centering
			\includegraphics[width=28mm,height=25mm]{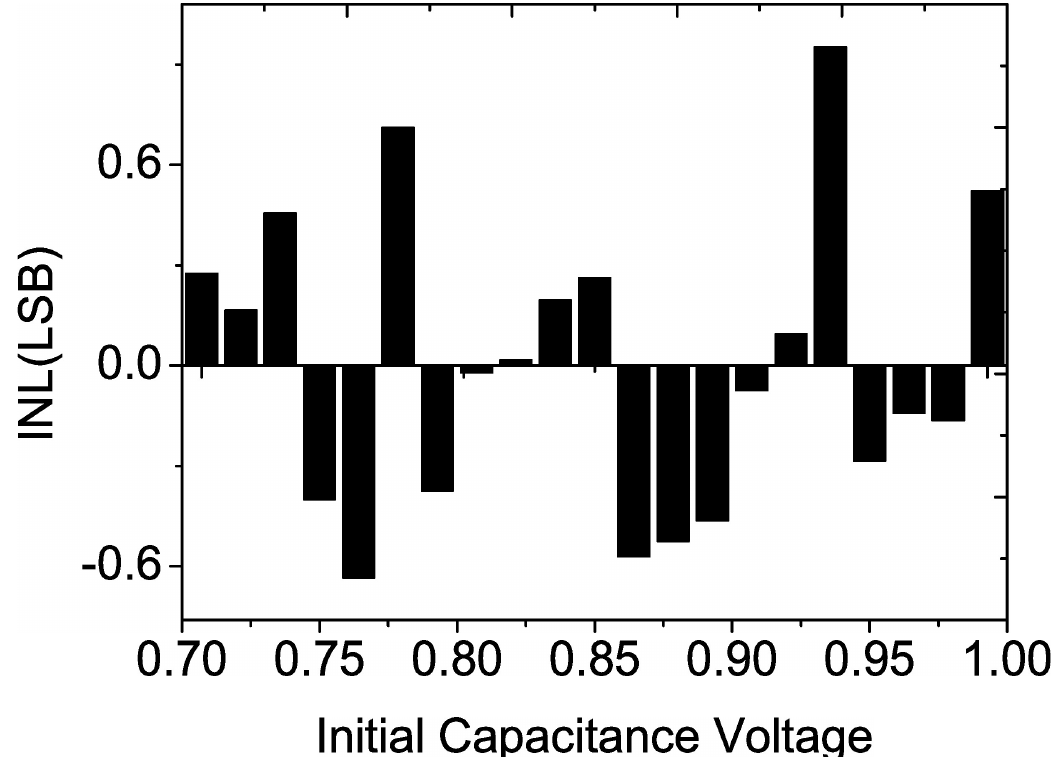}
		\end{minipage}
		\begin{minipage}[]{0.3\linewidth}
			\centering
			\includegraphics[width=28mm,height=25mm]{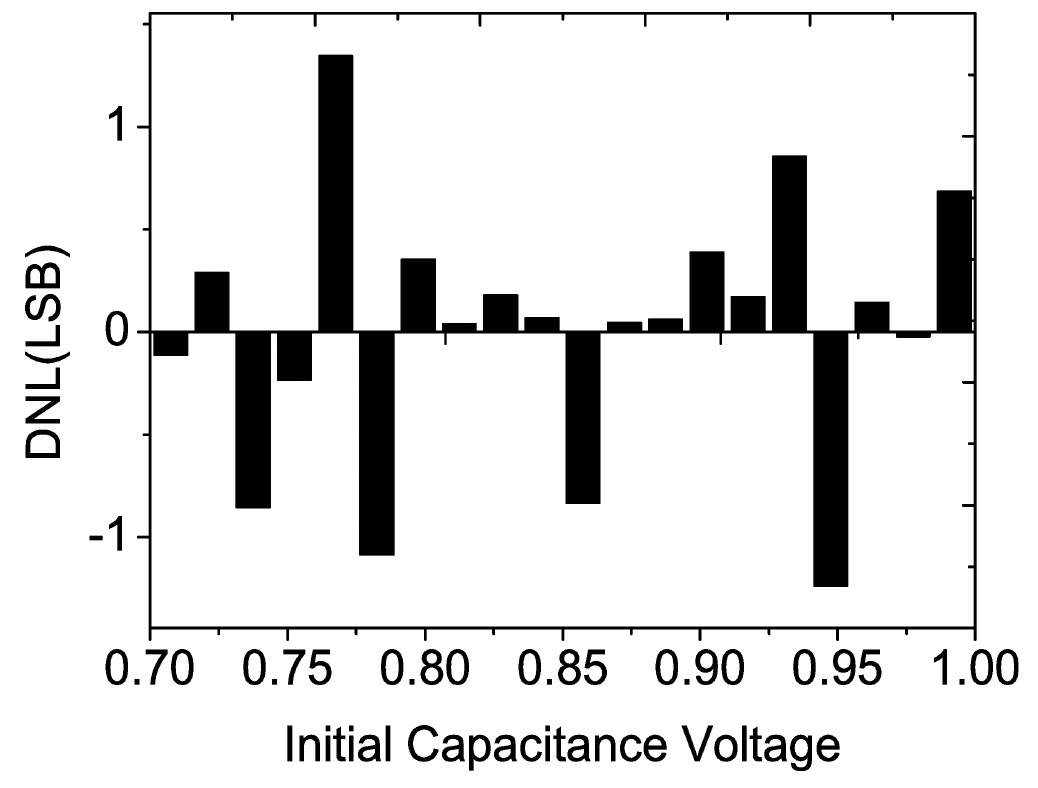}
		\end{minipage}
	}
	\subfigure[]{\label{with_T0}
		\begin{minipage}[]{0.3\linewidth}
			\centering
			\includegraphics[width=28mm,height=25mm]{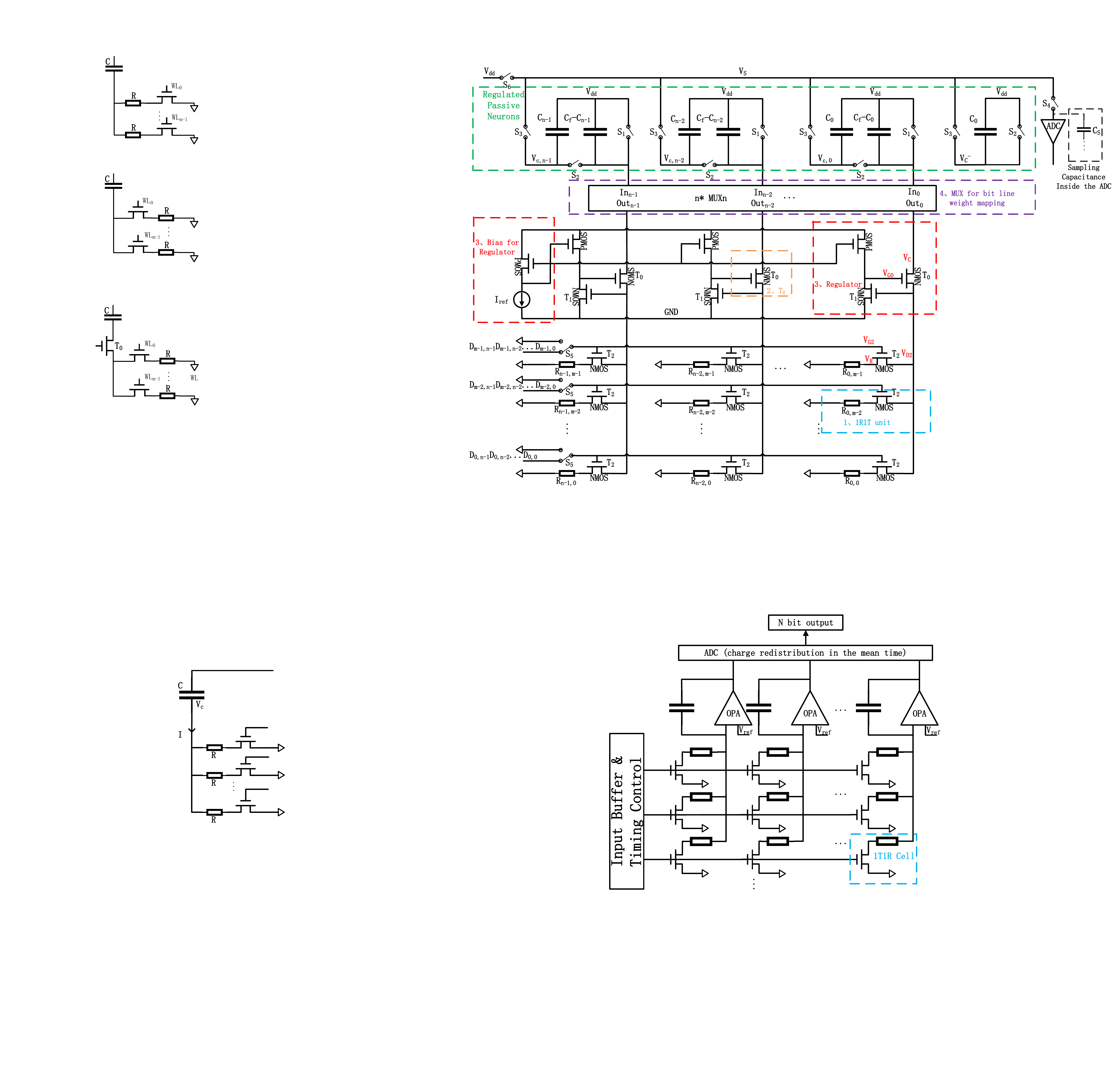}
		\end{minipage}
		\begin{minipage}[]{0.3\linewidth}
			\centering
			\includegraphics[width=28mm,height=25mm]{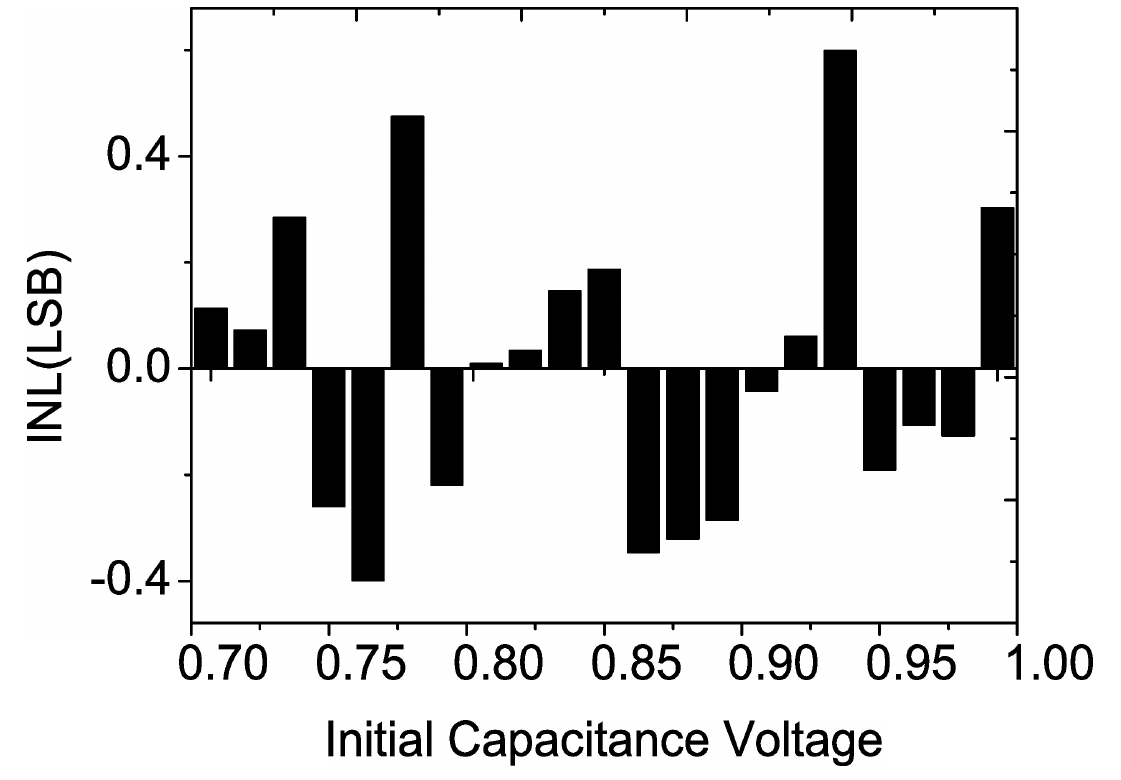}
		\end{minipage}
		\begin{minipage}[]{0.3\linewidth}
			\centering
			\includegraphics[width=28mm,height=25mm]{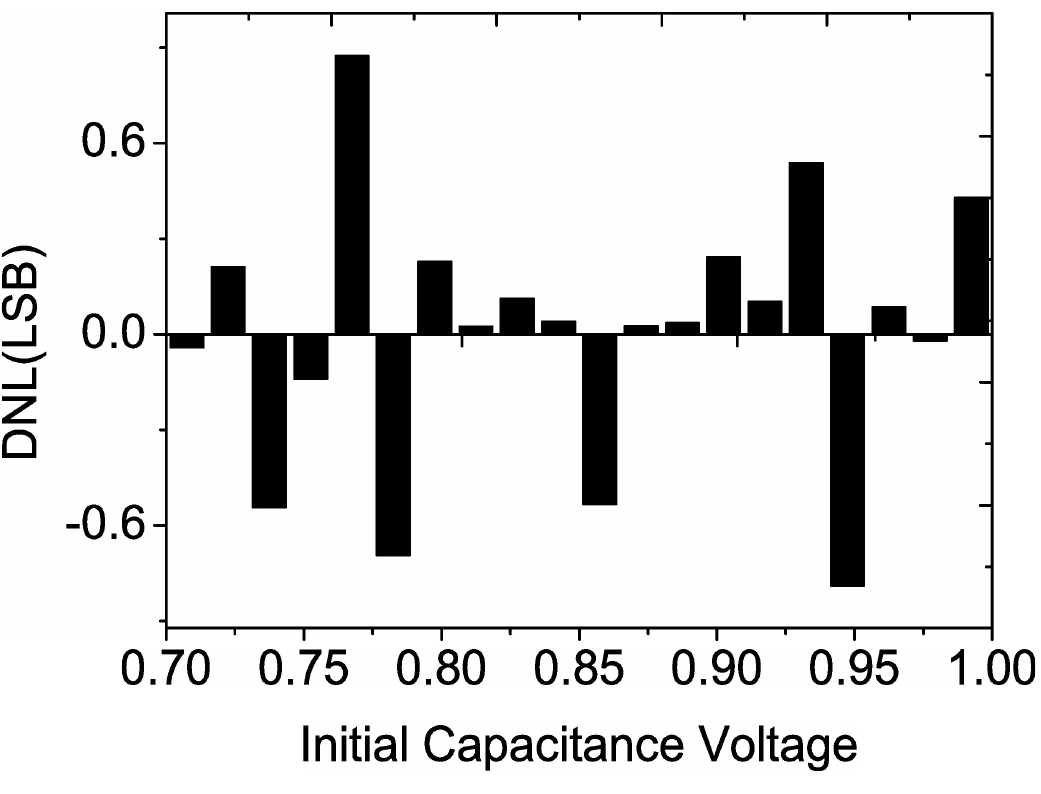}
		\end{minipage}
	}
	
	\caption{The INL/DNL comparison of integration results under different integrating voltage(0.7$\sim$1V) between \subref{1R1T} integrator without 1T1R unit position switching, \subref{no_T0} integrator without T0, and \subref{with_T0} integrator with $T_0$.}
\end{figure}

\begin{figure}[]
	
	
	
	\subfigure[]{
		\label{input}
		\begin{minipage}[]{0.5\linewidth}
			\centering
			\includegraphics[width=40mm,height=30mm]{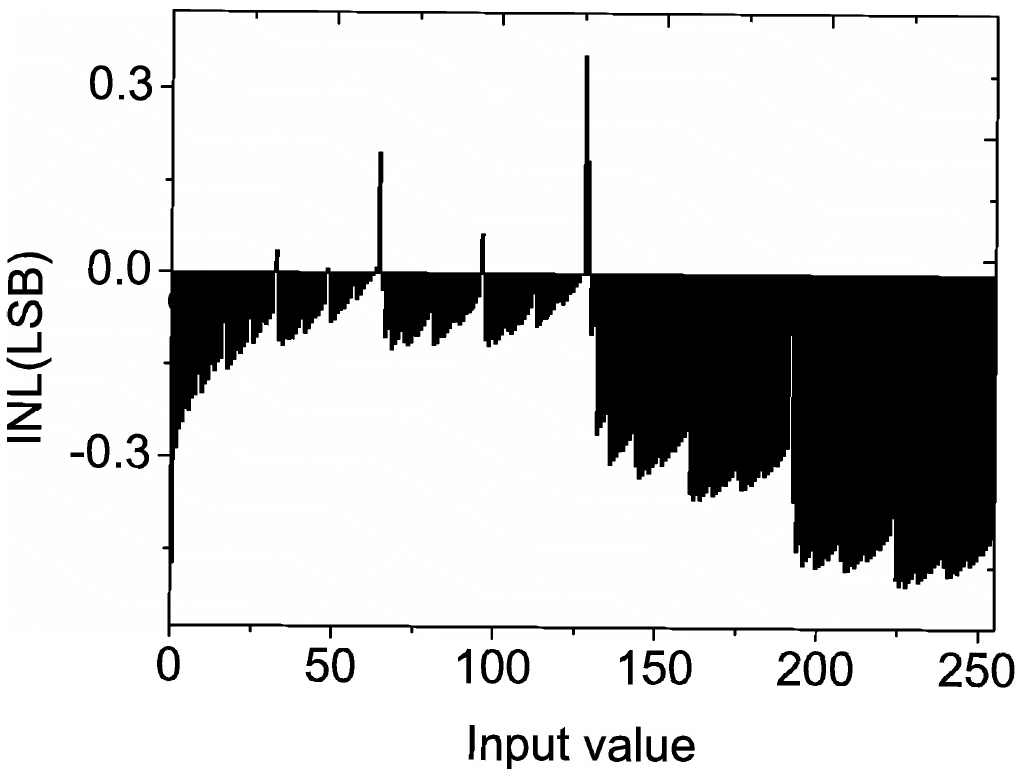}\\
		\end{minipage}
		\begin{minipage}[]{0.5\linewidth}
			\centering
			\includegraphics[width=40mm,height=30mm]{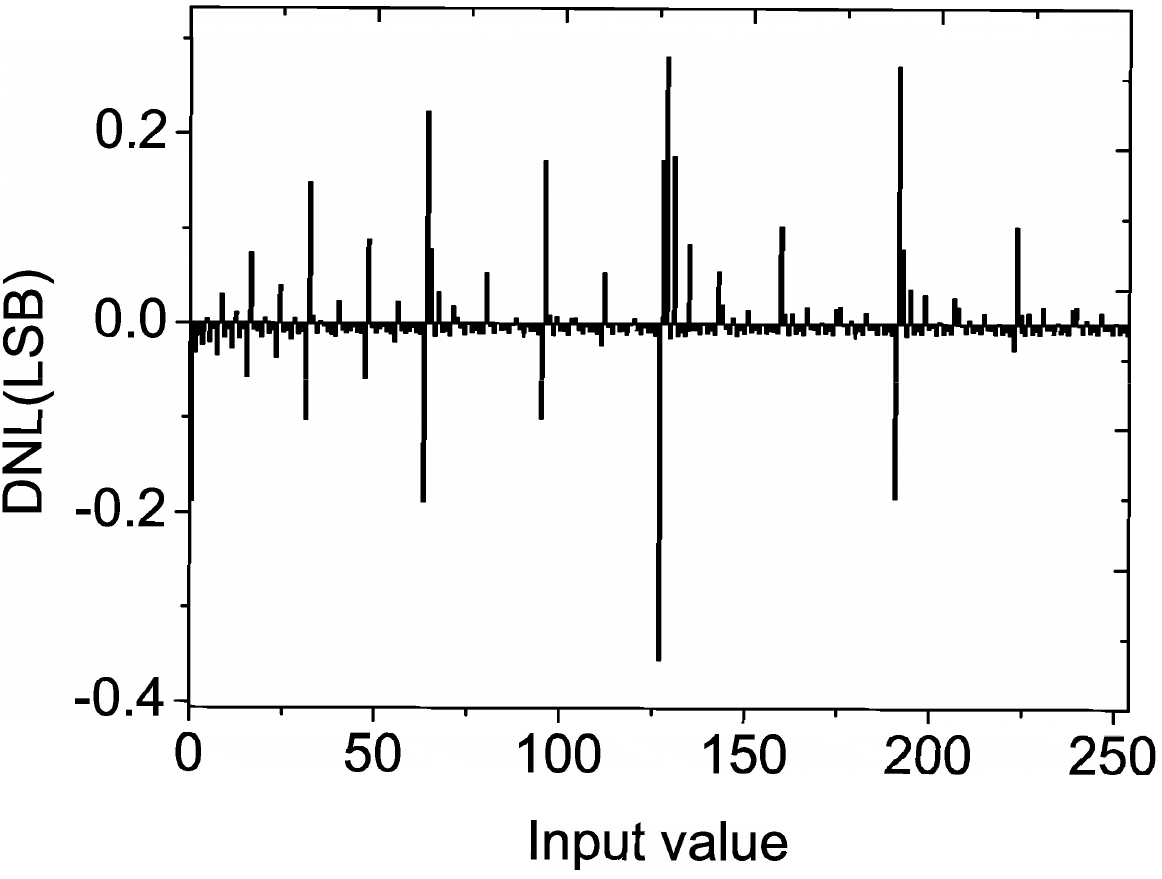}
		\end{minipage}
	}
	\subfigure[]{\label{weight}
		\begin{minipage}[]{0.5\linewidth}
			\centering
			\includegraphics[width=40mm,height=30mm]{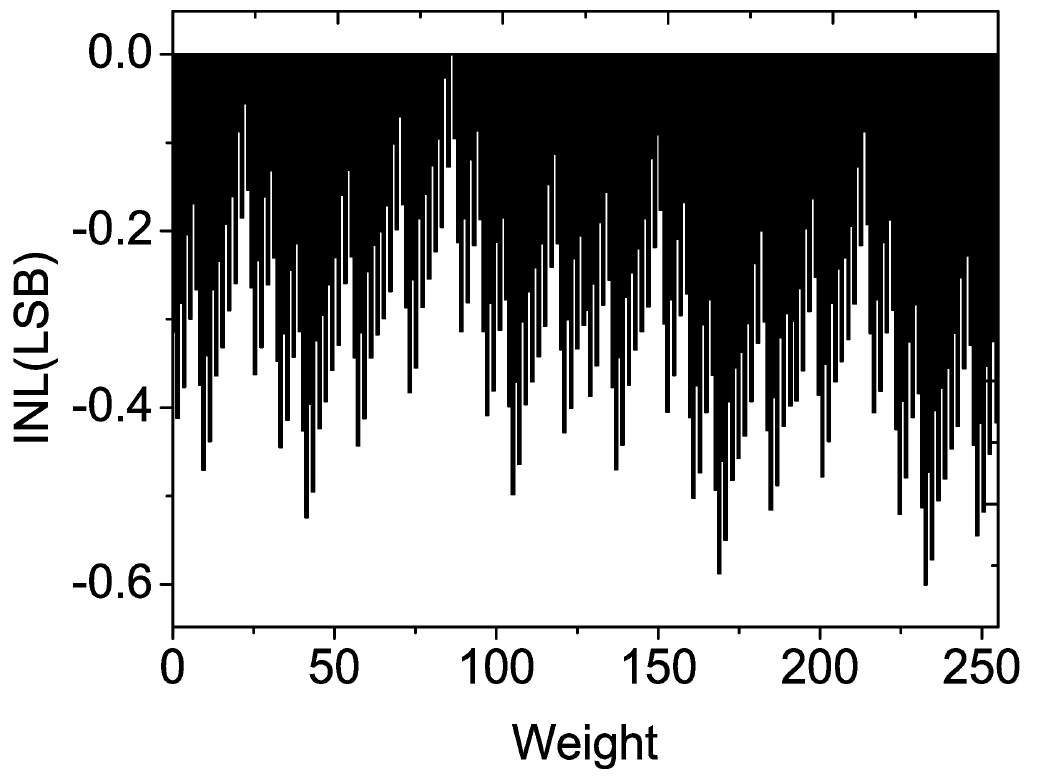}\\
		\end{minipage}
		\begin{minipage}[]{0.5\linewidth}
			\centering
			\includegraphics[width=40mm,height=30mm]{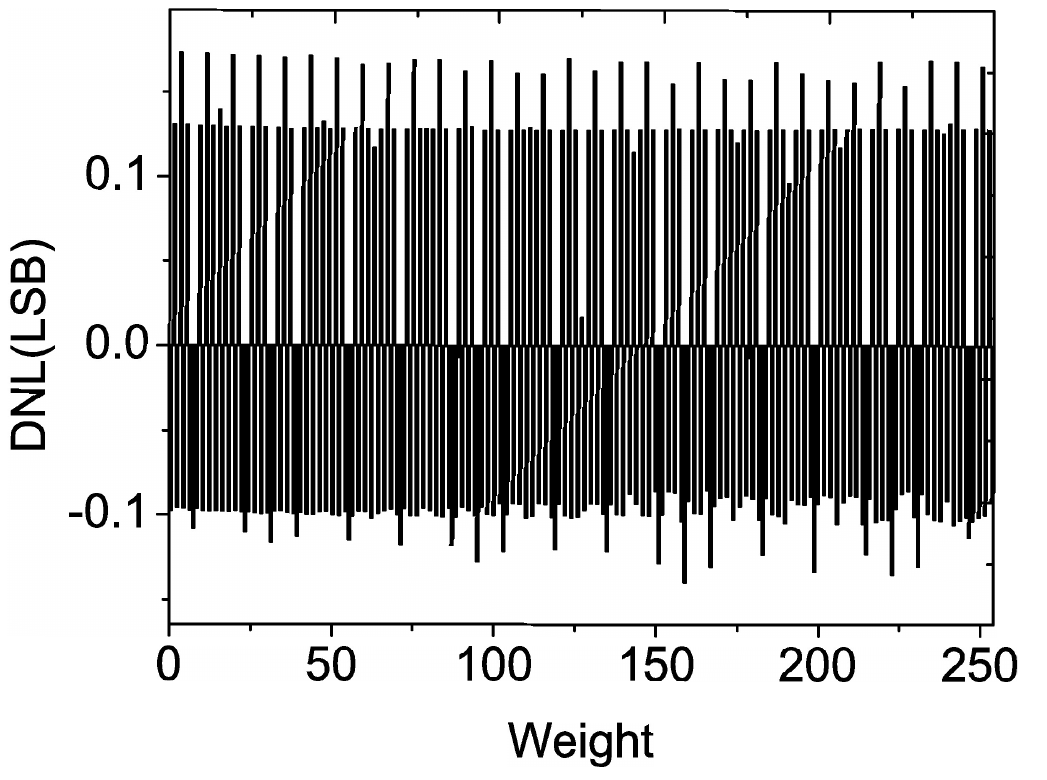}
		\end{minipage}
	}
	\subfigure[]{\label{regulator_inputlines}
		\begin{minipage}[]{0.5\linewidth}
			\centering
			\includegraphics[width=40mm,height=30mm]{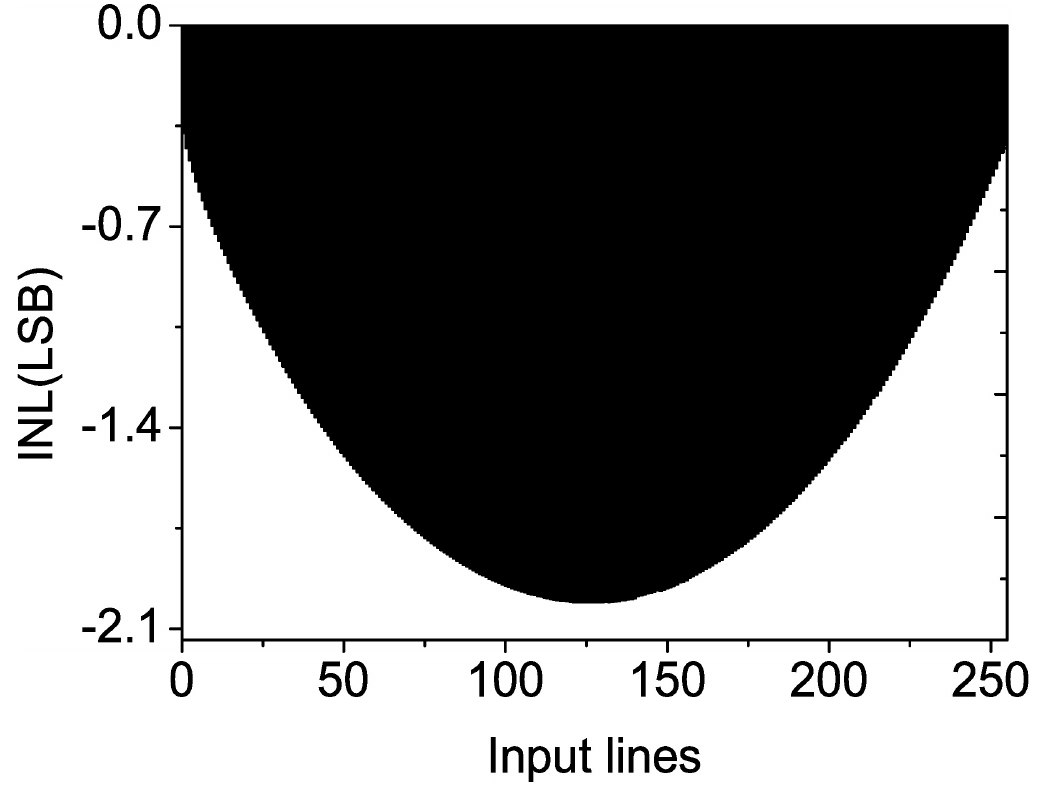}
		\end{minipage}
		\begin{minipage}[]{0.5\linewidth}
			\centering
			\includegraphics[width=40mm,height=30mm]{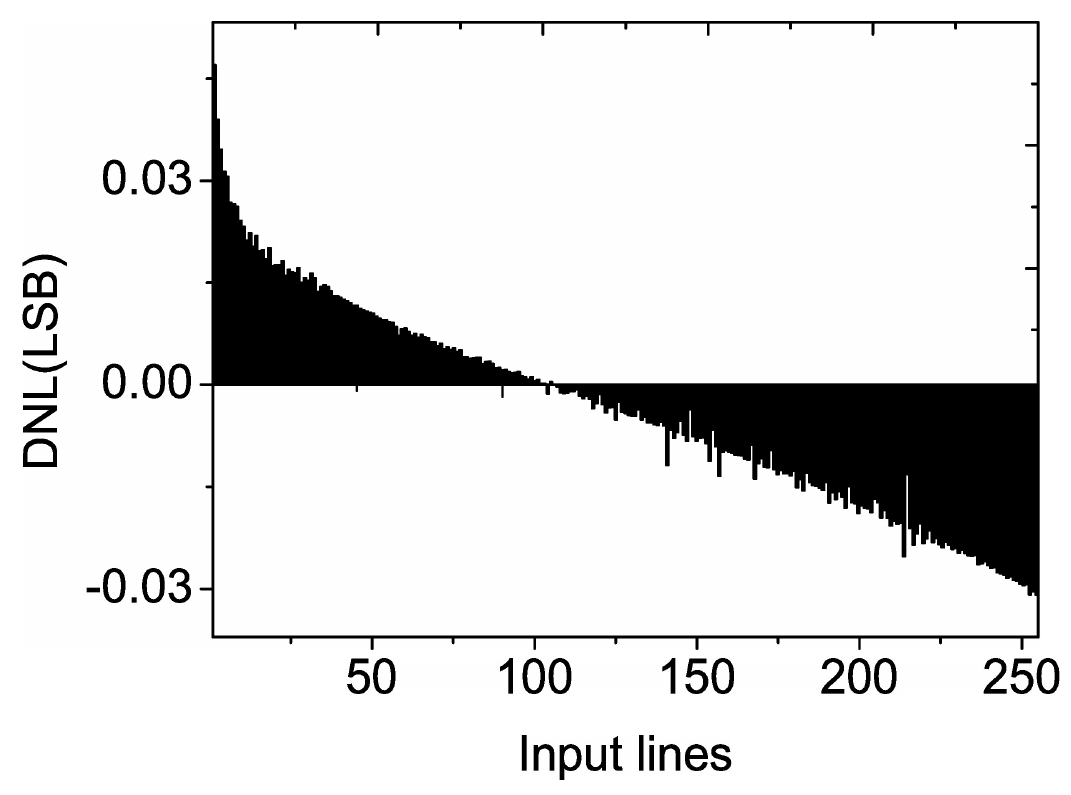}
		\end{minipage}
	}
	\subfigure[]{\label{noreg_inputlines}
		\begin{minipage}[]{0.5\linewidth}
			\centering
			\includegraphics[width=40mm,height=30mm]{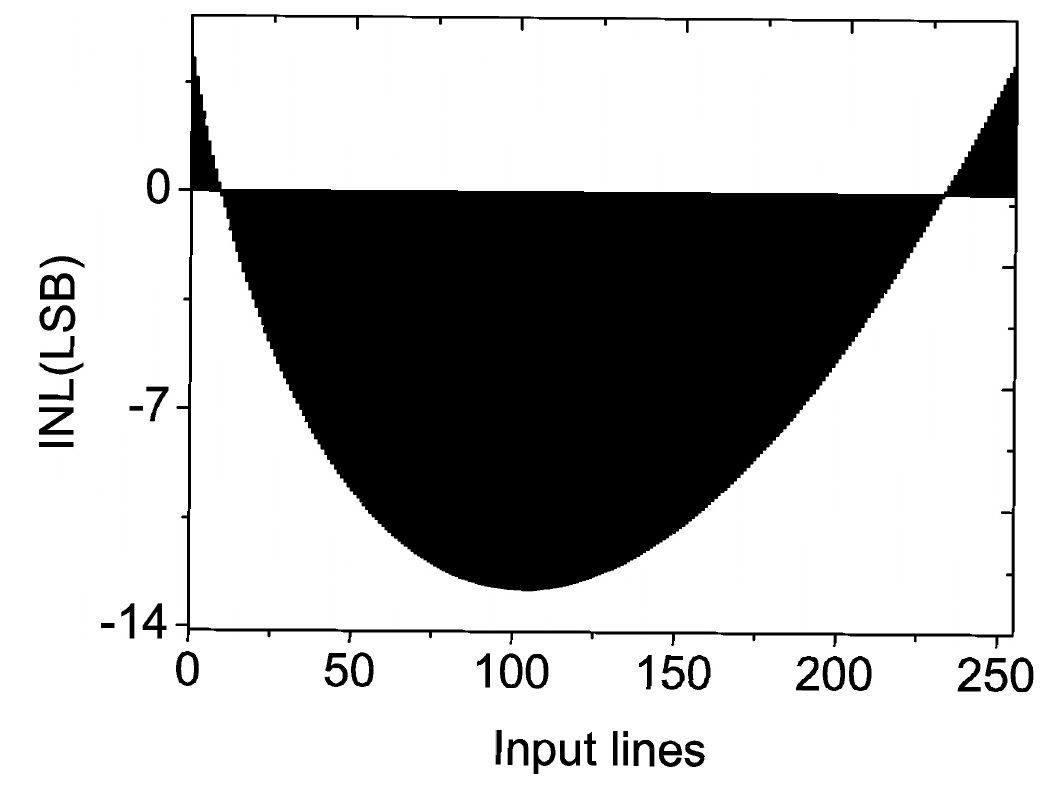}
		\end{minipage}
		\begin{minipage}[]{0.5\linewidth}
			\centering
			\includegraphics[width=40mm,height=30mm]{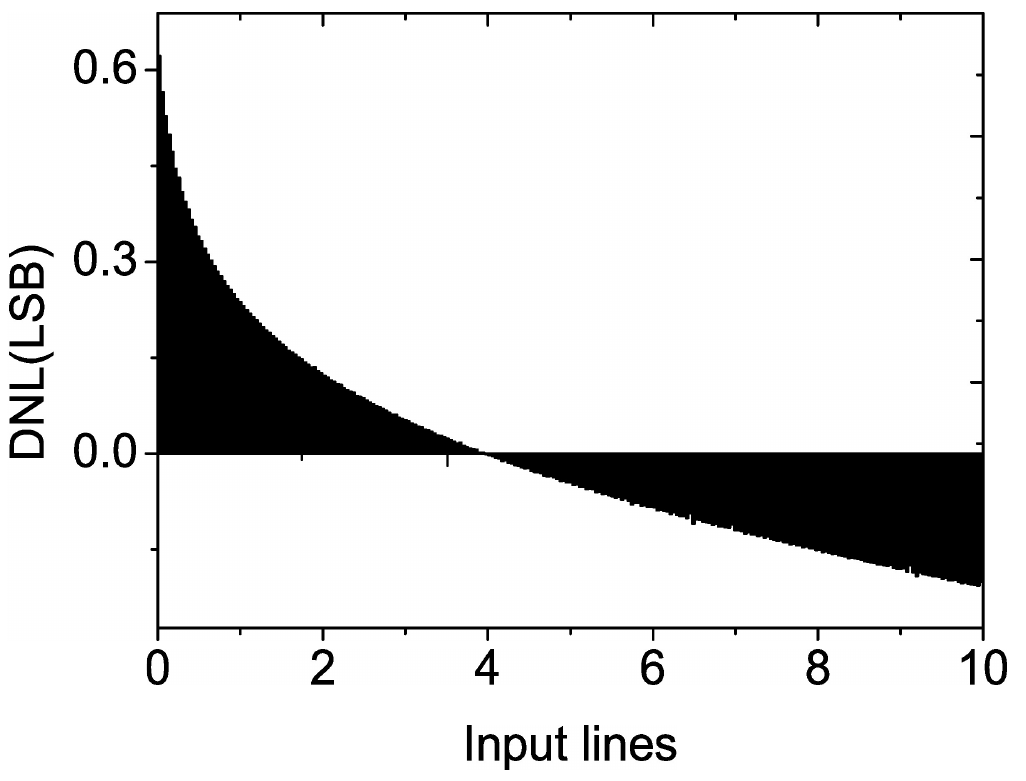}
		\end{minipage}
	}
	\caption{ The evaluation of linearity in terms of \subref{input} different input(0-255) and \subref{weight} different weight(0-255) and the INL/DNL comparison between \subref{regulator_inputlines} integral multiplier with regulator and \subref{noreg_inputlines} integral multiplier without regulator under different input lines(1-256).}
	\label{linearity simulation}
\end{figure}

The linearity comparison of integration results under different initial integrating voltage (0.7$\sim$1V) between the integrator without 1T1R unit position switching, integrator without $T_0$, and integrator with $T_0$ is shown in Fig. \ref{1R1T}, Fig. \ref{no_T0}, and Fig. \ref{with_T0}, respectively. The Differential Nonlinearity (DNL) and Integration Nonlinearity (INL) are used to evaluate the performance. The INL/DNL is (-1.66$\sim$0.89)/(-2.19$\sim$1.95) LSB for the integrator without 1T1R unit position switching, (-0.63$\sim$0.95)/(-1.24$\sim$1.35) LSB for integrator without $T_0$, and (-0.40$\sim$0.60)/(-0.79$\sim$0.87) LSB for the integrator with $T_0$ which confirms that the linearity of the integration process is greatly improved by 1T1R unit position switching and $T_0$. Fig. \ref{input} and Fig. \ref{weight} present the linearity evaluation of the proposed integral multiplier with different input and weight by the code density measurement. The circuit achieves INL/DNL of (-0.51$\sim$0.36)/(-0.35$\sim$0.28) LSB, (-0.60$\sim$0.001)/(-0.14$\sim$0.17) LSB corresponding to input value and weight, respectively. The linearity comparison of the integral multiplier under different input lines between the circuit with regulator and without regulator is shown in Fig. \ref{regulator_inputlines} and Fig. \ref{noreg_inputlines}, respectively. The INL/DNL is (-2.01$\sim$-0.38)/(0.01$\sim$0.01) LSB for the circuit with regulator and (-12.7$\sim$4.26)/(-0.2$\sim$0.62) for the circuit without regulator, which shows that the linearity in terms of the number of input lines is significantly improved by the regulator for providing a relatively stable drain voltage of 1R1T when the loads of bit line change.\par
\begin{table}[h]
	\caption{PVT Simulation on ENOB}
	\resizebox{88mm}{6mm}{
		\begin{tabular}{|l|l|l|l|l|l|l|l|l|l|}
			\hline
			Process & \multicolumn{4}{l|}{ff} & \multicolumn{4}{l|}{ss} & tt \\ \hline
			Temperature($^{\circ}$C) & \multicolumn{2}{l|}{-40} & \multicolumn{2}{l|}{80} & \multicolumn{2}{l|}{-40} & \multicolumn{2}{l|}{80} & 27 \\ \hline
			Voltage(V) & 0.9 & 1.1 & 0.9 & 1.1 & 0.9 & 1.1 & 0.9 & 1.1 & 1 \\ \hline
			ENOB(bit) & 7.36 & 7.3 & 7.25 & 7.1 & 7.35 & 7.27 & 7.05 & 7.03 & 7.37 \\ \hline
	\end{tabular}}
	\label{PVT}
\end{table}
\begin{figure}[t]
	\centering
	\includegraphics[height=45mm]{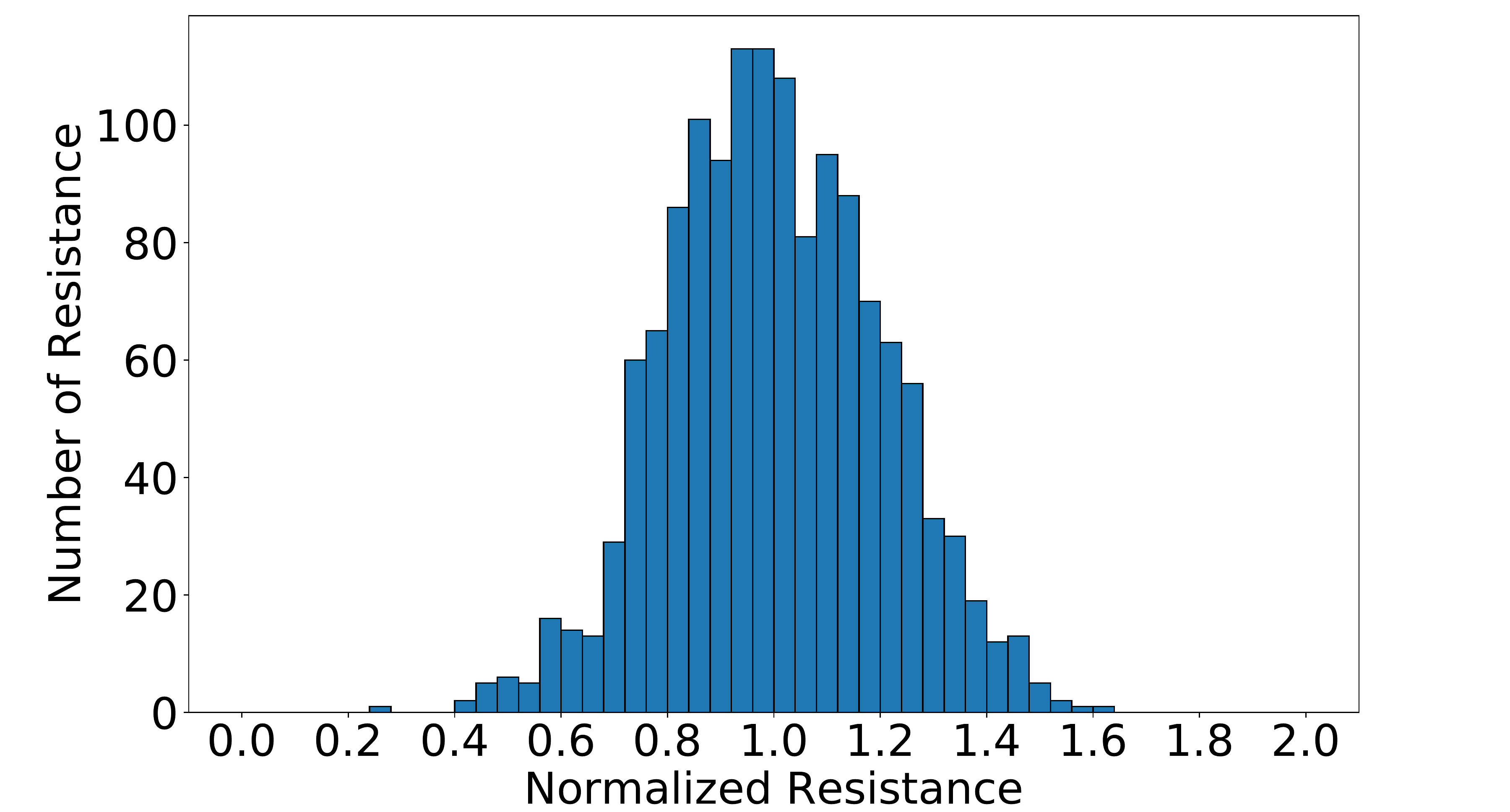}
	\caption{Normalized resistance distribution read by ADC where the standard deviation of normalized Gaussian distribution is 0.2.}
	\label{resistance distribution}
\end{figure}

\begin{figure}[]
	\subfigure[]{\label{normal mapping}
		\begin{minipage}[]{1\linewidth}
			\centering
			\includegraphics[height=50mm]{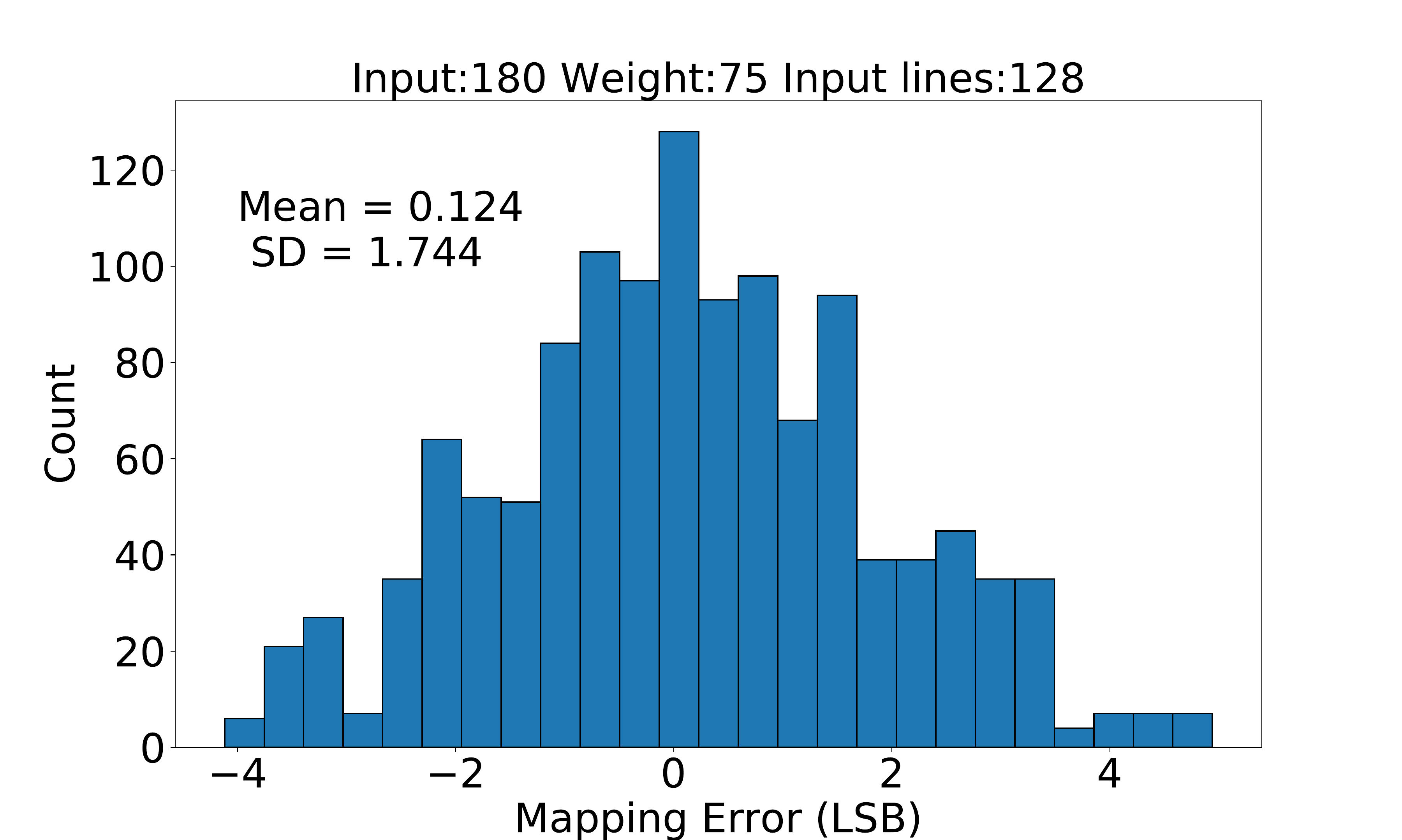}
	\end{minipage}}
	\subfigure[]{\label{bit line weight mapping}
		\begin{minipage}[]{1\linewidth}
			\centering
			\includegraphics[height=50mm]{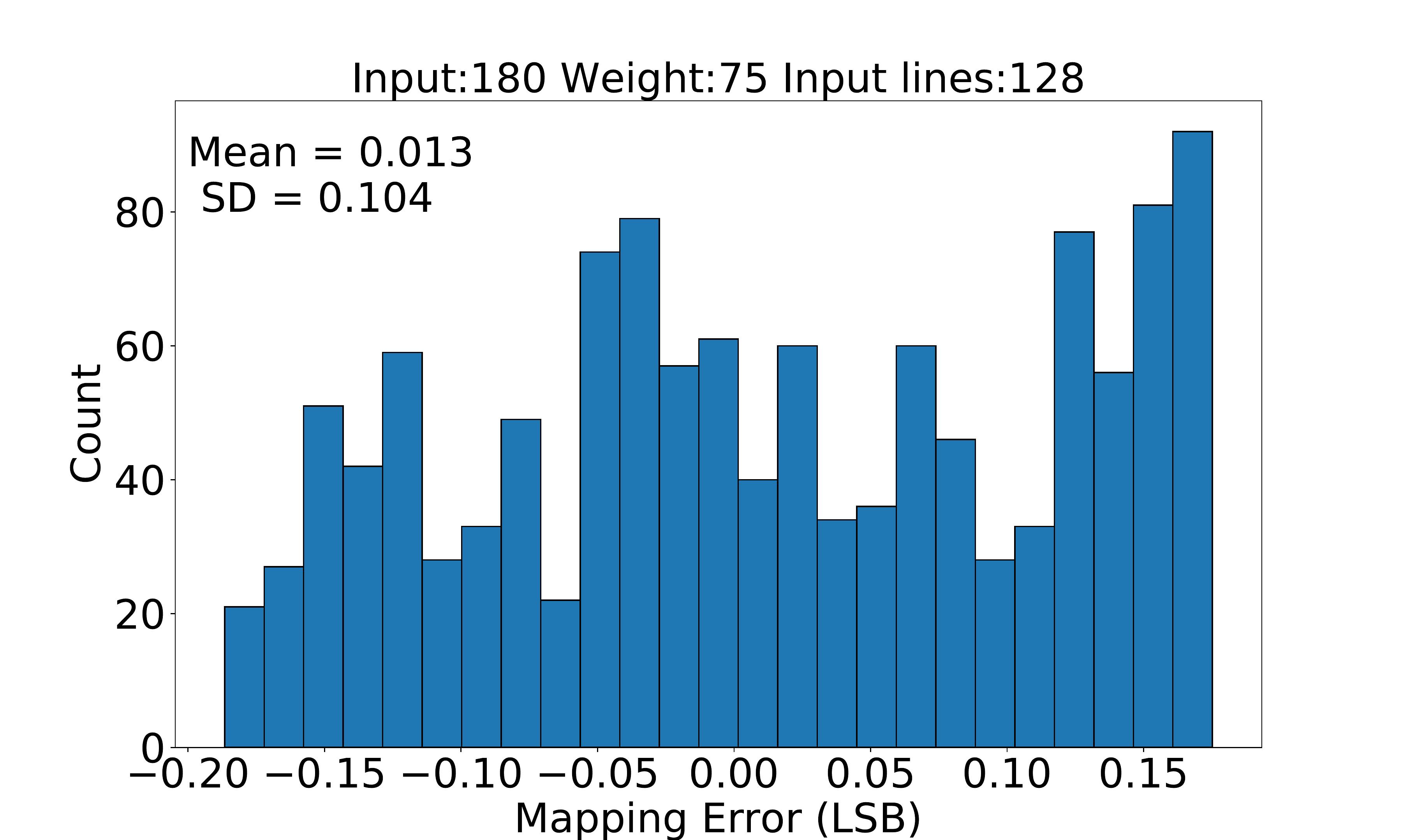}
		\end{minipage}
	}
	
	\caption{Comparison of 1400 Monte Carlo simulations on the computation error of combination of input 180, weight 75, number of input lines 128 between \subref{normal mapping} normal mapping and \subref{bit line weight mapping} bit line weight mapping method.}
	\label{Monte_mapping}
\end{figure}

\subsubsection{PVT Simulation}
To verify the robustness of the circuit, different combinations of process, voltage, and temperature are chosen to do the PVT simulation where ENOB is used to evaluate the core's performance. The ENOBs in these PVT combinations are all greater than 7 bits as shown in Table \ref{PVT} which indicates that the proposed circuit is reliable with variations of process, voltage, and temperature. 

\subsubsection{Quantization and Mapping Methods Comparison}
To add the impact of resistance distribution into the weight of the neural network, the resistance reading phase is needed when the ADC is used to read the resistance of the RRAM array. To make things easy, the process of ADC reading 1R1T circuit with fixed resistance is firstly simulated by 1400 Monte Carlo simulations to evaluate the impact of the transistor variation, then the resistance inconsistency is evaluated by adding a normalized Gaussian distribution. The normalized distribution of RRAM array read by ADC is shown in Fig.\ref{resistance distribution}, where the standard deviation of the normalized Gaussian distribution is 0.2. Fig. \ref{Monte_mapping} shows the comparison of 1400 Monte Carlo simulations on the computation error of combination of input 180, weight 75, number of input lines 128 between normal mapping and bit line weight mapping method. The average value and the standard deviation of the error in normal mapping method are 0.124 LSB and 1.744LSB, respectively, while those of the errors in bit line weight mapping method are 0.013 LSB and 0.104 LSB, respectively. The mapping result indicates that the bit line weight mapping method significantly improves our CIM core's robustness to variations of device inconsistency.\par
\begin{figure*}[]
	\subfigure[]{\label{AlexNet}
		\begin{minipage}[]{1\linewidth}
			\centering
			\includegraphics[width=85mm,height=37mm]{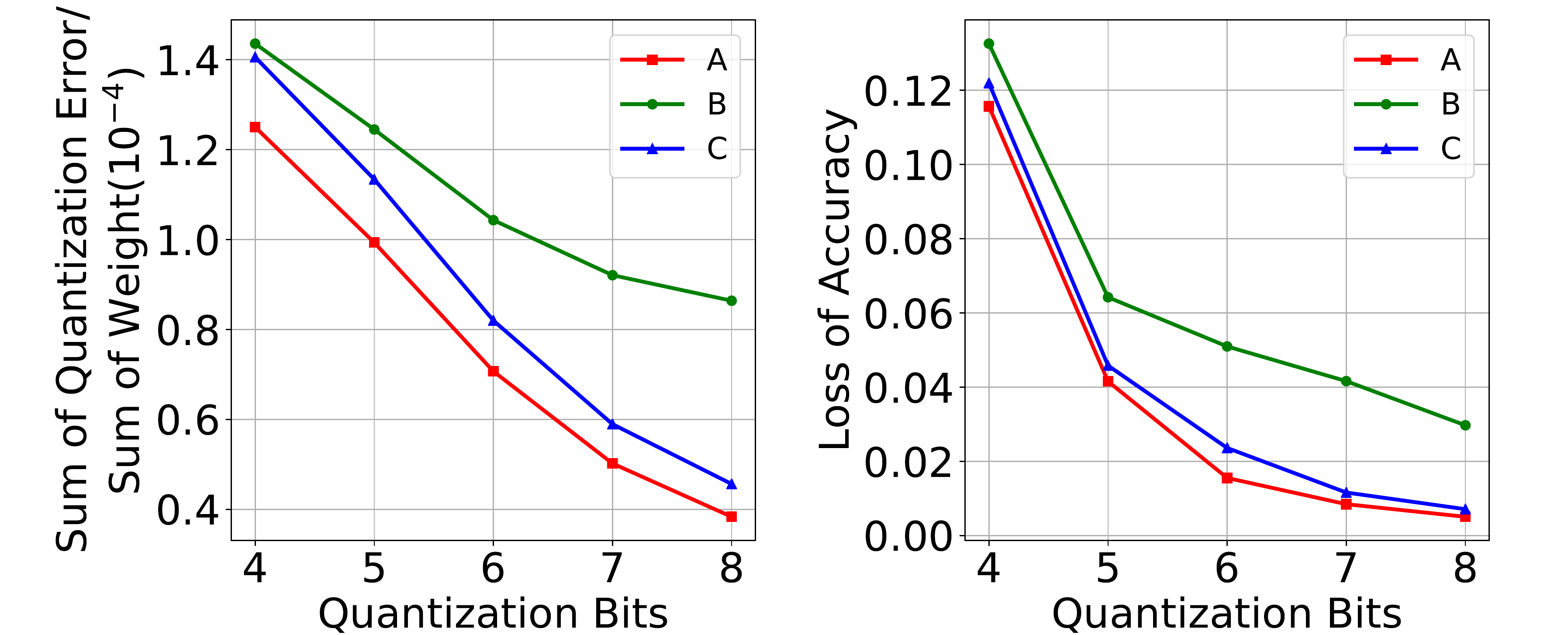}
			\centering
			\includegraphics[width=80mm]{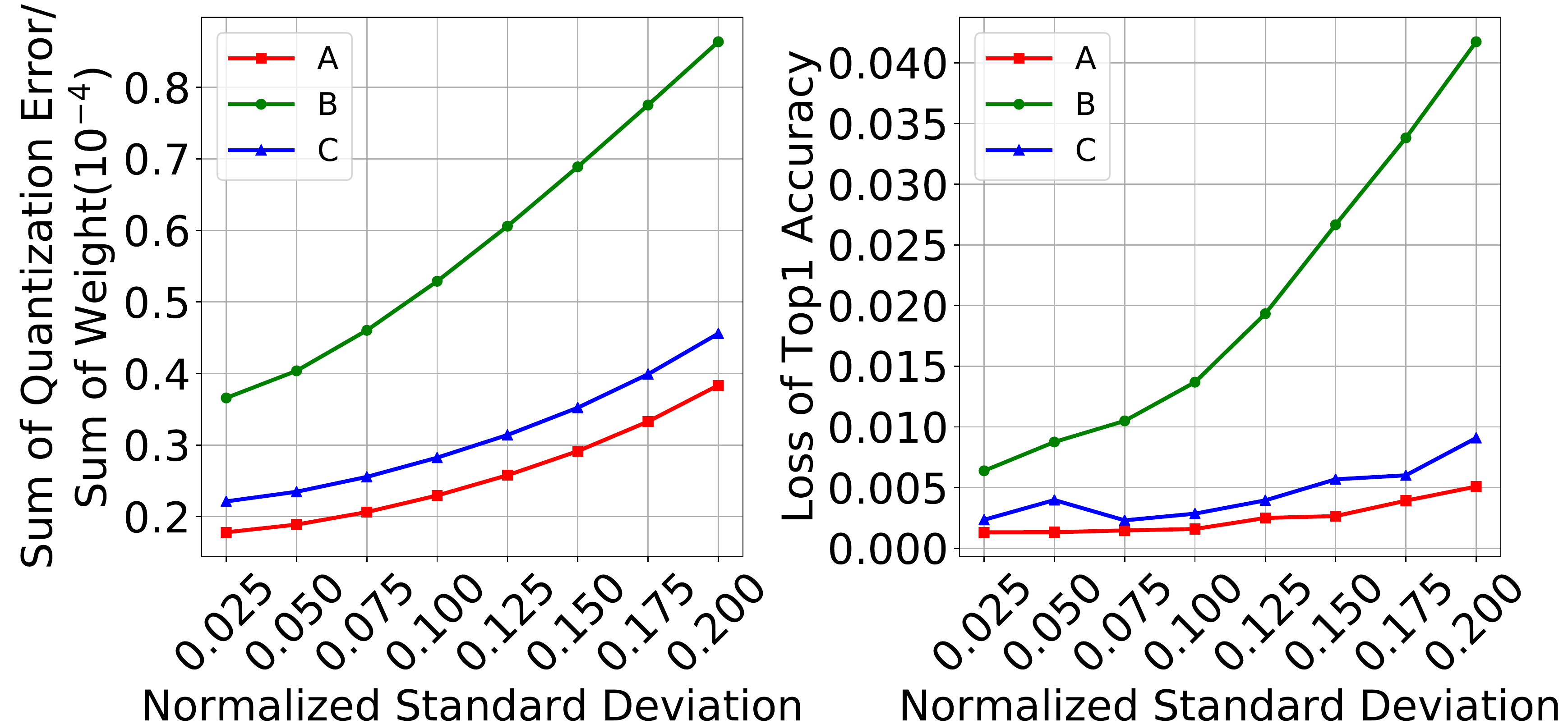}
		\end{minipage}
	}
	
	\subfigure[]{\label{VGG16}
		\begin{minipage}[]{1\linewidth}
			\centering
			\includegraphics[width=85mm,height=37mm]{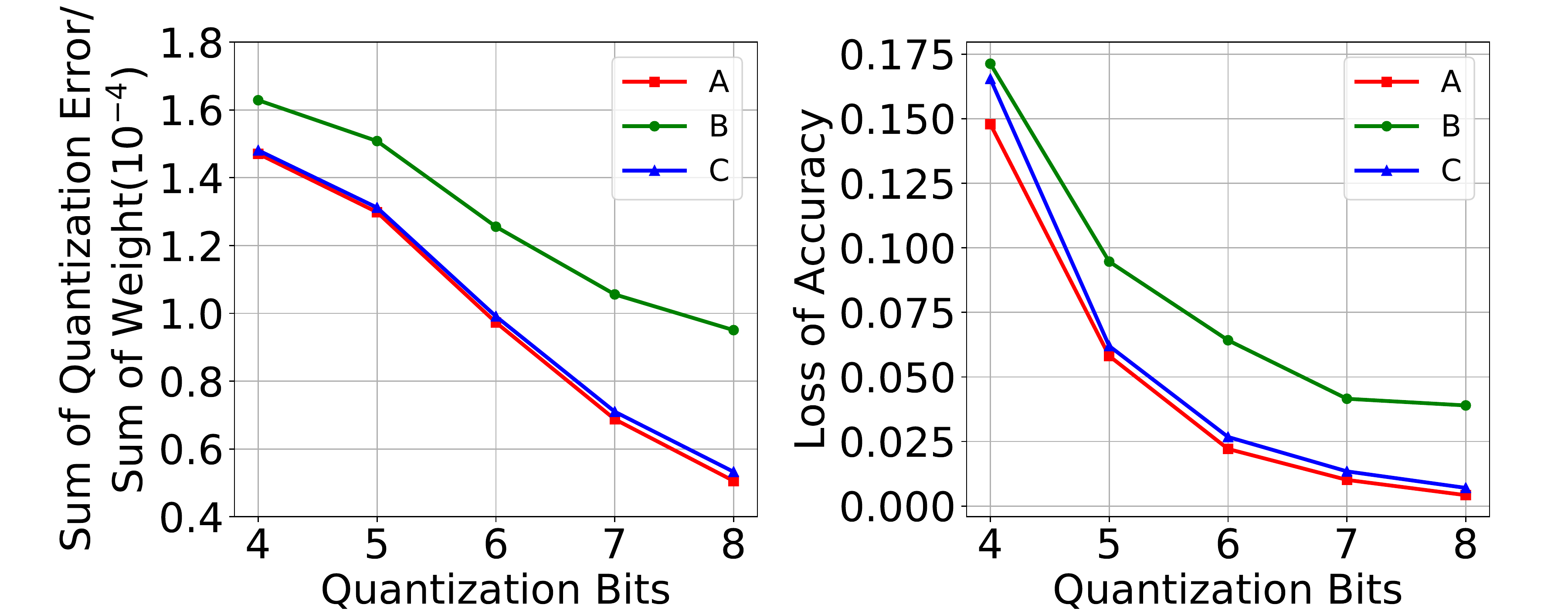}
			\centering
			\includegraphics[width=80mm]{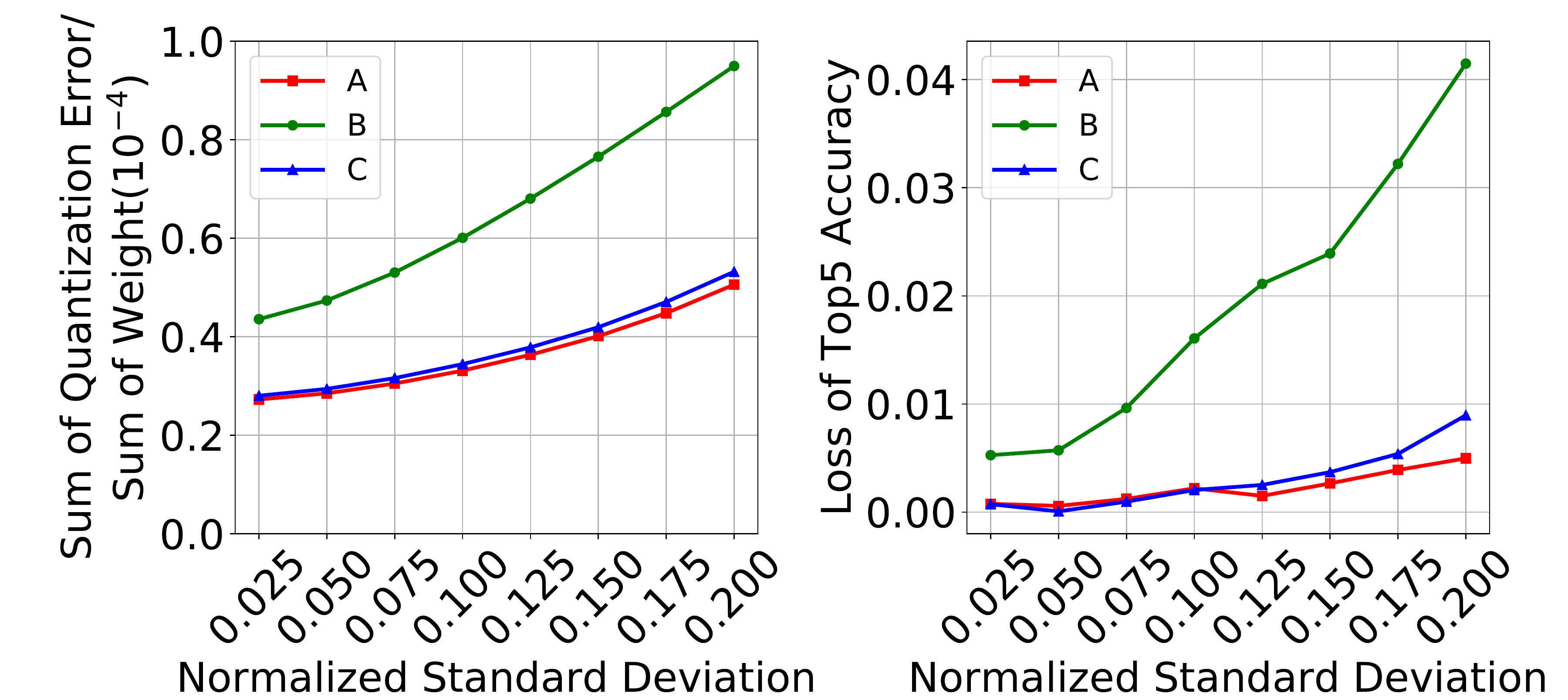}
		\end{minipage}
	}

	\caption{The accuracy comparison between A: Resistance based quantization and bit line weight mapping method, B: Normal binary quantization and mapping method, and C: Resistance based quantization and normal mapping method. \subref{AlexNet} The left two figures are quantization error ratio (sum of absolute values of quantization error/sum of absolute values of weight) and loss of top-1 accuracy of AlexNet on ILSVRC 2012 with different quantization bits (the deviation of normalized resistance is 0.2) while the right two are quantization error ratio and loss of top-1 accuracy of AlexNet on ILSVRC 2012 with different standard deviation of normalized resistance distribution (the quantization bits is 8). \subref{VGG16} The left two figures are quantization error ratio and loss of top-1 accuracy of VGG16 on ILSVRC 2012 with different quantization bits (the deviation of normalized resistance is 0.2) while the right two are quantization error ratio and loss of top-1 accuracy of VGG16 on ILSVRC 2012 with different standard deviation of normalized resistance distribution (the quantization bits is 8).}
	\label{quan_comparison}
\end{figure*}

To test the effect of the bit line weight mapping method on network level, three quantization and mapping methods are simulated. The first one is normal binary quantization and mapping method, which quantifies the weight to digital 8-bit value and set the resistance HRS/LRS according to the corresponding digital bit 0/1. The second one is resistance based quantization and mapping method that quantifies the weight according to Eq. \ref{quantization equ}. The third one is resistance based quantization and bit line weight mapping method proposed in this paper. Fig. \ref{AlexNet} shows the quantization error ratio (sum of absolute values of quantization error/sum of absolute values of weight) and the loss of top-1 accuracy comparison between three methods under different quantization bits (the deviation of normalized resistance is 0.2) and different standard deviation of normalized resistance distribution (the quantization bits is 8) on AlexNet and ILSVRC 2012; Fig. \ref{VGG16} presents the quantization error ratio and the loss of top-1 accuracy comparison between quantization methods with different quantization bit (the deviation of normalized resistance is 0.2) and standard deviation of the resistance distribution (the quantization bits is 8) on VGG16 and ILSVRC 2012. As shown in Fig. \ref{quan_comparison}, the optimized quantization and bit line weight mapping method helps reduce the quantization errors and improve the inference accuracy both on AlexNet and VGG16. For example, the accuracy loss in 8-bit mode with 0.2 deviation on AlexNet are 2.97\% and 0.51\% for normal binary mapping method and bit line weight mapping method, respectively, and those on VGG16 are 2.70\% and 0.39\% for normal binary mapping method and bit line weight mapping method, respectively. What's more, with the uncertainty of the resistance increasing the effect of the optimization is more evident.\par

\section{Conclusion}
In this paper, an 8-bit RRAM based CIM core with regulated passive neuron and bit line weight mapping method has been proposed. The non-linearity brought by the passive integrator and the errors caused by quantization and the cell to cell variation have been discussed. To address the above issues, the detailed regulated integral multiplier and the bit line weight mapping method have been presented. The circuit level simulation has shown that the proposed CIM core achieves 3.61mW on power consumption with the size of 256*256 in 8-bit input and 8-bit weight mode, which is reduced by 98.2\% compared with MBRAI while the SFDR and SNDR of the CIM core achieve 59.13 dB and 46.13 dB, respectively. The network level simulation has shown that the CIM core achieves 0.90\% top-1 error rate with 0.013 uJ/img on LeNet and 43.60\% top-1 error rate with 16.65 uJ/img on AlexNet, which are better than other schemes. The linearity and PVT simulation has been done to verify the robustness of the circuit. The simulation on mapping methods has shown that compared with normal mapping method, the proposed bit line weight mapping scheme achieves better performance which improves the top-1 accuracy by 2.46\% and 3.47\% for AlexNet and VGG16 on ILSVRC 2012 in 8-bit mode.

\section*{Acknowledgment}

This work was supported by the Major Scientic Research Project of Zhejiang Lab (No. 2019KC0AD02).

\ifCLASSOPTIONcaptionsoff
  \newpage
\fi



%


\bibliographystyle{unsrt.bst}
\bibliography{ref.bib}
%

\begin{IEEEbiography}[{\includegraphics[width=1in,height=1.25in,clip,keepaspectratio]{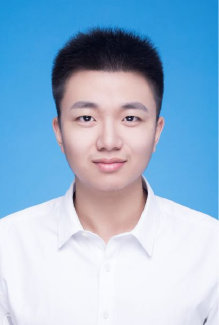}}]{Yewei Zhang}
(Student Member, IEEE) recieved the bachelor’s degree from College of Information Science \& Electronic Engineering, Zhe Jiang University in 2018. He is currently studying for a master's degree at College of Information Science \& Electronic Engineering, Zhe Jiang University. He is interested in in-memory computing and non-volatile memories.
\end{IEEEbiography}

\begin{IEEEbiography}[{\includegraphics[width=1in,height=1.25in,clip,keepaspectratio]{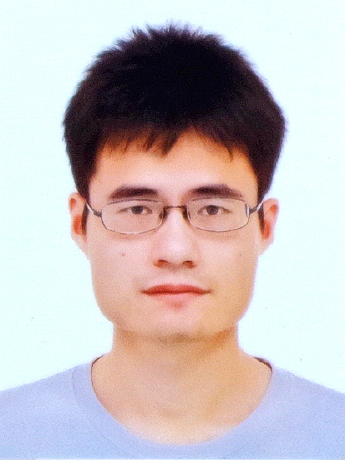}}]{Kejie Huang}
	(Senior Member, IEEE) received
	the Ph.D. degree from the Department of Electrical
	Engineering, National University of Singapore
	(NUS), Singapore, in 2014. He has been a Principal
	Investigator with the College of Information Science
	Electronic Engineering, Zhejiang University (ZJU),
	since 2016. Prior to joining ZJU, he has spent five
	years at the IC design industry, including Samsung
	and Xilinx, two years in the Data Storage Institute,
	Agency for Science Technology and Research
	(A*STAR), and another three years in the Singapore
	University of Technology and Design (SUTD), Singapore. He has authored
	or coauthored more than 30 scientific articles in international peer-reviewed
	journals and conference proceedings. He holds four granted international
	patents, and another eight pending ones. His research interests include low
	power circuits and systems design using emerging non-volatile memories,
	architecture and circuit optimization for reconfigurable computing systems
	and neuromorphic systems, machine learning, and deep learning chip design.
	He currently serves as the Associate Editor of the IEEE TRANSACTIONS ON
	CIRCUITS AND SYSTEMS-PART II: EXPRESS BRIEFS.
\end{IEEEbiography}

\begin{IEEEbiography}[{\includegraphics[width=1in,height=1.25in,clip,keepaspectratio]{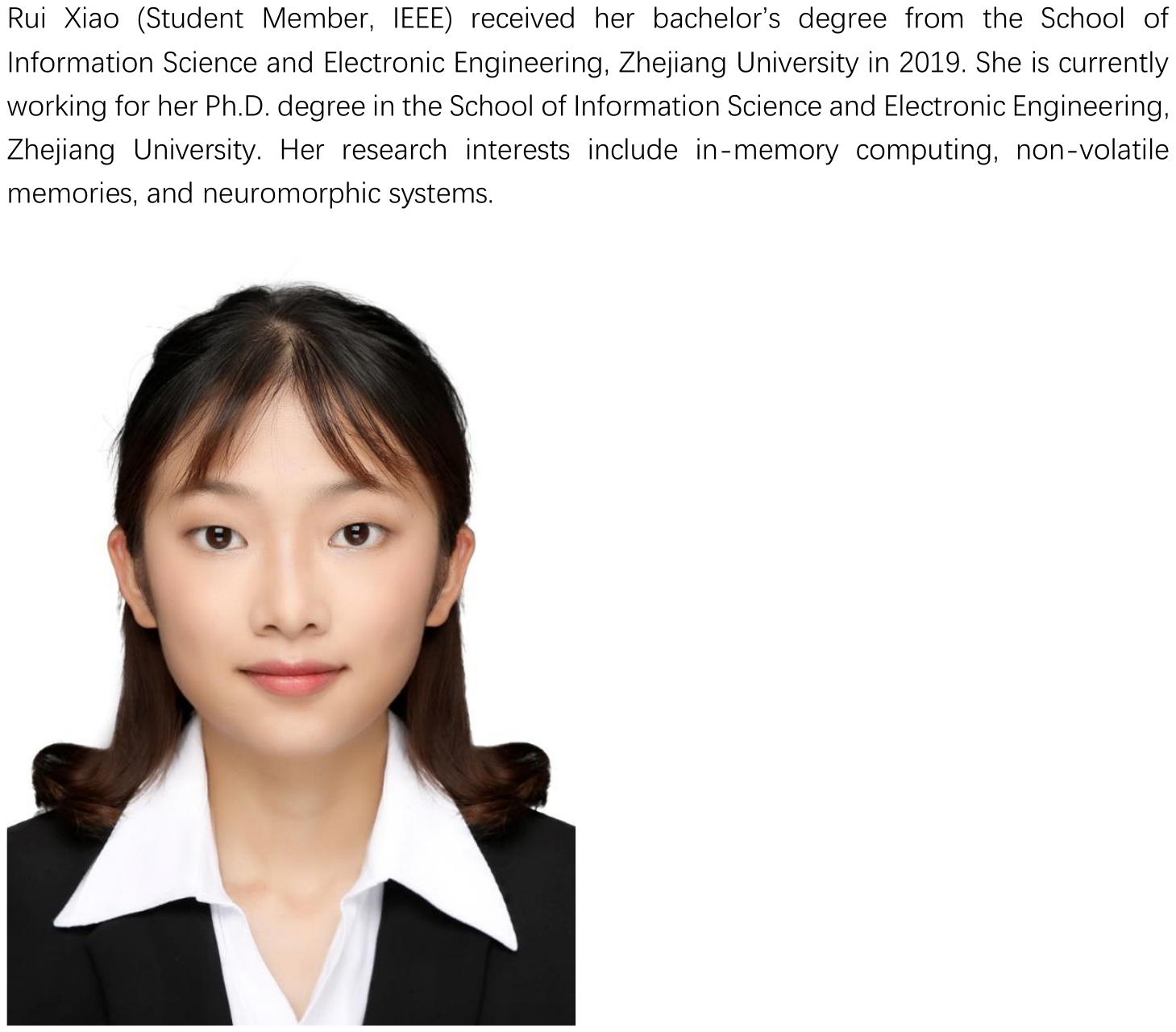}}]{RuiXiao}
(Student Member, IEEE) received her bachelor’s degree from the School of Information Science and Electronic Engineering, Zhejiang University in 2019. She is currently working for her Ph.D. degree in the School of Information Science and Electronic Engineering, Zhejiang University. Her research interests include in-memory computing, non-volatile memories, and neuromorphic systems.
\end{IEEEbiography}


\begin{IEEEbiography}[{\includegraphics[width=1in,height=1.25in,clip,keepaspectratio]{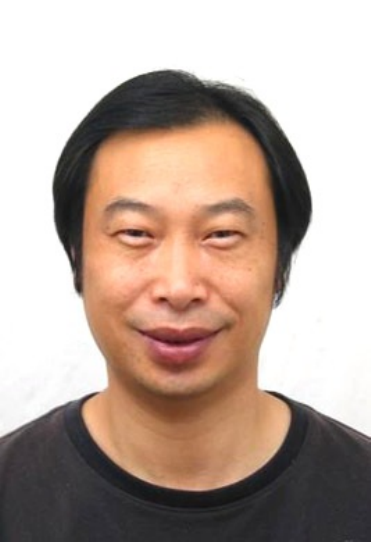}}]{Haibin Shen}
is currently a Professor with Zhejiang
University, a member of the second level of 151 talents
project of Zhejiang Province, and a member
of the Key Team of Zhejiang Science and Technology
Innovation. His research interests include learning
algorithm, processor architecture, and modeling.
His research achievement has been used by many
authority organizations. He has published more than
100 papers on academic journals, and he has been
granted more than 30 patents of invention. He was a
recipient of the First Prize of Electronic Information
Science and Technology Award from the Chinese Institute of Electronics, and
has won a second prize at the provincial level.
\end{IEEEbiography}








\end{document}